\documentclass[sigconf, nonacm]{acmart}

\usepackage{float}

\newcommand\vldbdoi{XX.XX/XXX.XX}
\newcommand\vldbpages{XXX-XXX}
\newcommand\vldbvolume{14}
\newcommand\vldbissue{1}
\newcommand\vldbyear{2020}
\newcommand\vldbauthors{\authors}
\newcommand\vldbtitle{\shorttitle} 
\newcommand\vldbavailabilityurl{https://github.com/orepapas/What_Blocks_My_Blockchains_Throughput_Data}
\newcommand\vldbpagestyle{plain} 

\usepackage{enumitem}
\usepackage{orcidlink}
\usepackage{algorithmic}
\usepackage{graphicx}
\usepackage{textcomp}
\usepackage{xcolor}
\usepackage{xurl}
\usepackage{url}
\usepackage{xcolor}
\usepackage[nolist]{acronym}
\usepackage{subfig}
\usepackage{balance}

\begin{document}
\title{What Blocks My Blockchain’s Throughput? \\ Developing a Generalizable Approach for Identifying Bottlenecks in Permissioned Blockchains}

\newif\ifdraft
\drafttrue
\ifdraft
\newcommand{\jsnote}[1]{ {\textcolor{red} { ***Johannes: #1 }}}
\newcommand{\opnote}[1]{ {\textcolor{orange} { ***Orestis: #1 }}}
\newcommand{\lbnote}[1]{ {\textcolor{purple} { ***Lasse: #1 }}}
\newcommand{\eenote}[1]{ {\textcolor{cyan} { ***Egor: #1 }}}
\newcommand{\GFapproved}[1]{\leavevmode\color{lightgray}{}#1}
\newcommand{\GFignore}[1]{\renewcommand{\acp}{}}
\else
\newcommand{\jsnote}[1]{}
\newcommand{\opnote}[1]{}
\newcommand{\lbnote}[1]{}
\fi

\author{Orestis Papageorgiou}
\orcid{0000-0003-3412-5082}
\affiliation{%
  \institution{SnT -- Interdisciplinary Centre for Security, Reliability and Trust}
  \institution{University of Luxembourg}
  \city{Luxembourg}
  \state{Luxembourg}
}
\email{orestis.papageorgiou@uni.lu}

\author{Lasse Börtzler}
\affiliation{%
  \institution{Dept. of Economics and Management}
  \institution{Karlsruhe Institute of Technology}
  \city{Karlsruhe}
  \country{Germany}
}
\email{lasse.boertzler@student.kit.edu}

\author{Egor Ermolaev}
\orcid{0000-0003-3412-5082}
\affiliation{%
  \institution{SnT -- Interdisciplinary Centre for Security, Reliability and Trust}
  \institution{University of Luxembourg}
  \city{Luxembourg}
  \country{Luxembourg}
}
\email{egor.ermolaev@uni.lu}

\author{Jyoti Kumari}
\affiliation{%
  \institution{SnT -- Interdisciplinary Centre for Security, Reliability and Trust}
  \institution{University of Luxembourg}
  \city{Luxembourg}
  \country{Luxembourg}
}
\email{jyoti.kumari@uni.lu }

\author{Johannes Sedlmeir}
\orcid{0000-0003-2631-8749}
\affiliation{%
  \institution{SnT -- Interdisciplinary Centre for Security, Reliability and Trust}
  \institution{University of Luxembourg}
  \city{Luxembourg}
  \country{Luxembourg}
}
\email{johannes.sedlmeir@uni.lu}

\begin{abstract}
Permissioned blockchains have been proposed for a variety of use cases that require decentralization yet address enterprise requirements that permissionless blockchains to date cannot satisfy -- particularly in terms of performance. However, popular permissioned blockchains still exhibit a relatively low maximum throughput in comparison to established centralized systems. Consequently, researchers have conducted several benchmarking studies on different permissioned blockchains to identify their limitations and -- in some cases -- their bottlenecks in an attempt to find avenues for improvement. Yet, these approaches are highly heterogeneous, difficult to compare, and require a high level of expertise in the implementation of the underlying specific blockchain. In this paper, we develop a more unified and graphical approach for identifying bottlenecks in permissioned blockchains based on a systematic review of related work, experiments with the Distributed Ledger Performance Scan (DLPS), and an extension of its graphical evaluation functionalities. We conduct in-depth case studies on Hyperledger Fabric and Quorum, two widely used permissioned blockchains with distinct architectural designs, demonstrating the adaptability of our framework across different blockchains. We provide researchers and practitioners working on evaluating or improving permissioned blockchains with a toolkit, guidelines on what data to document, and insights on how to proceed in the search process for bottlenecks.
\end{abstract}

\maketitle

\pagestyle{\vldbpagestyle}
\begingroup\small\noindent\raggedright\textbf{PVLDB Reference Format:}\\
\vldbauthors. \vldbtitle. PVLDB, \vldbvolume(\vldbissue): \vldbpages, \vldbyear.\\
\href{https://doi.org/\vldbdoi}{doi:\vldbdoi}
\endgroup
\begingroup
\renewcommand\thefootnote{}\footnote{\noindent
This work is licensed under the Creative Commons BY-NC-ND 4.0 International License. Visit \url{https://creativecommons.org/licenses/by-nc-nd/4.0/} to view a copy of this license. For any use beyond those covered by this license, obtain permission by emailing \href{mailto:info@vldb.org}{info@vldb.org}. Copyright is held by the owner/author(s). Publication rights licensed to the VLDB Endowment. \\
\raggedright Proceedings of the VLDB Endowment, Vol. \vldbvolume, No. \vldbissue\ %
ISSN 2150-8097. \\
\href{https://doi.org/\vldbdoi}{doi:\vldbdoi} \\
}\addtocounter{footnote}{-1}\endgroup

\ifdefempty{\vldbavailabilityurl}{}{
\vspace{.3cm}
\begingroup\small\noindent\raggedright\textbf{PVLDB Artifact Availability:}\\
The source code, data, and/or other artifacts have been made available at \url{\vldbavailabilityurl}.
\endgroup
}

\begin{acronym}
\acro{iot}[IoT]{Internet of Things}
\acro{dlps}[DLPS]{distributed ledger performance  scan}
\acro{CPU}{central processing unit}
\acro{tps}{transactions per second}
\acro{freq}[f$_{\text{req}}$]{request frequency}
\acro{fresp}[f$_{\text{resp}}$]{response frequency}
\acro{vscc}[VSCC]{\textit{validation system chaincode}}
\acro{mvcc}[MVCC]{\textit{multi version concurrency control}}
\acro{eda}[EDA]{\textit{exploratory data analysis}}
\acro{kde}[KDE]{kernel density estimate}
\acro{2pc}[2PC]{two-phase commit}
\acro{pow}[PoW]{proof of work}
\acro{pos}[PoS]{proof of stake}
\acro{raft}[RAFT]{Reliable, Replicated, Redundant, And Fault-Tolerant}
\acro{ibft}[IBFT]{Istanbul Byzantine Fault Tolerant}
\acro{qbft}[QBFT]{Quorum Byzantine Fault Tolerant}
\acro{evm}[EVM]{Ethereum Virtual Machine}
\acro{dapp}[DApp]{decentralized application}
\acro{api}[API]{application programming interface}
\acro{dlt}[DLT]{distributed ledger technology}
\acrodefplural{dlt}[DLTs]{distributed ledger technologies}
\acro{tx}[TX]{transaction}
\acrodefplural{tx}[TXs]{transactions}
\acro{rpc}[RPC]{remote procedure call}
\acrodefplural{rpc}[RPCs]{remote procedure calls}
\end{acronym}
\section{Introduction}
\label{sec:introduction}

Since its inception by Nakamoto in 2008~\cite{nakamoto_bitcoin_2008}, blockchain technology has been explored across various industries far beyond its original use in decentralized payment systems. Researchers and practitioners have explored its potential in a variety of applications, including its use as a database~\cite{peng_falcondb_2020, nathan_blockchain_2019, ge_hybrid_2022, hong_gridb_2023}, in the~\ac{iot}~\cite{dai2019blockchain, tseng_blockchain-based_2020, lockl2020blockchainIoT}, in cross-organizational workflow \citep{fridgen2018crossorganizational}, supply chain management~\cite{guggenberger2020improving}, and in several other applications where a neutral platform is desirable~\cite{sedlmeir2022transparency}. In this context, organizations that aimed to use~\ac{dlt}-based systems also explored permissioned blockchains that restrict not only data visibility but also participation in consensus. This approach removes the limitations in terms of storage, bandwidth, and computing that permissionless systems impose to maintain a high degree of decentralization. Moreover, permissioned blockchains use different consensus mechanisms for replicated state machines~\cite{vukolic2015quest}, for instance, PBFT~\cite{castro2002practical}, which allow for lower latency compared to common consensus algorithms used in permissionless systems, such as~\ac{pow}, and~\ac{pos}. 

However, the transition from experimental blockchain projects to practical applications in business operations faces challenges. Many projects fail to advance beyond the pilot stage due to technical complexities~\cite{toufaily2021framework}. This has resulted in a slower adoption rate compared to other technologies developed around the same time~\cite{gartner_large}. One of the primary reasons behind this is the technology's inferior performance when compared to that of centralized systems. The resource-intensive nature of replication and consensus in blockchains hinders efficient scaling and parallelization, causing centralized architectures to outperform by far even permissioned blockchains~\cite{sedlmeir2022serverless}. As a result, a substantial part of computer science research on blockchain technology focuses on performance characteristics of permissioned blockchains~\cite{fan_performance_2020}.

Blockchain benchmarking research primarily focuses on high-level performance indicators, such as overall throughput and average latency. This is very useful when comparing different blockchain types~\cite{dinh_blockbench_2017} or when analyzing the effect of deployment parameters like the number of nodes, network delays, block size, and transaction type~\cite{androulaki_hyperledger_2018, 8845168, thakkar_performance_2018, kuzlu_performance_2019, guggenberger_-depth_2022}. Hardware utilization or networking-related aspects are frequently relegated to the background, included as aggregated data like maximum values, or considered as a configuration parameter~\cite{dinh_blockbench_2017,fan_performance_2020}. Additionally, blockchain benchmarking research remains heavily fragmented, blockchain-specific, and requires a high level of expertise to exercise.

This paper seeks to address these shortcomings by analyzing the behavior of hardware components within blockchain nodes, their interrelationships, and their impact on the network’s throughput in a systematic and graphical way, providing a general and illustrative method to detect bottlenecks. To achieve this, we examine Hyperledger Fabric (Fabric) and Quorum, as research has already provided many valuable insights and seem to be the most widely studied examples (see Section~\ref{sec:related}). We survey related work that we use to ground our method and conduct and evaluate our own experiments, using and extending the~\ac{dlps}, an open-source blockchain benchmarking framework that allows for the easy deployment and benchmarking of a broad range of blockchains and already supports limited graphical analyses~\cite{sedlmeir_dlps_2021}. We also draw parallels between our results and those of other researchers in an attempt to determine the accuracy of our method and findings.

The remainder of this paper is structured as follows: Section~\ref{sec:background} covers the fundamental characteristics of Fabric, Quorum, and the~\ac{dlps}. Section~\ref{sec:related} presents and structures related work in the field and points out the research gap to which we aim to contribute. Section~\ref{sec:m&f} then describes our bottleneck identification method and the corresponding analyses of Fabric and Quorum. We conclude the paper with a summary of our findings and discuss limitations and opportunities for future research in Section~\ref{sec:conclusion}.
\section{Background}
\label{sec:background}

\subsection{Hyperledger Fabric}
\label{fabric}

Fabric has risen to prominence as one of the industry's leading permissioned blockchains~\cite{zheng_nearly, guggenberger_-depth_2022}. Compared to other permissioned blockchains, its unique architecture provides novel opportunities for improving performance and eliminates the need for domain-specific and deterministic smart contract (called chaincode in the case of Fabric) programming languages~\cite{androulaki_hyperledger_2018}. 
Fabric's architecture relies on the \textit{execute-order-validate} paradigm. Under this approach, nodes first simulate (execute) transactions in any order, which are then batched and ordered in consensus. Finally, nodes validate the individual transactions and ensure that no conflicts have emerged owing to the potentially different ordering in the execution phase~\cite{androulaki_hyperledger_2018}. This approach is useful if cross-checking a computation-heavy transaction requires weaker agreement than an honest (super-)majority~\cite{guggenberger_-depth_2022}. At the ordering and/or validation stages, nodes can verify whether sufficient signatures for agreement, according to previously defined rules, are present. 

In Fabric, nodes are grouped into organizations, and a node can take at least one of the roles of a peer node (\emph{peer}) or an orderer (\emph{orderer})~\cite{androulaki_hyperledger_2018}. Only peers maintain an append-only ledger, whereas orderers create and broadcast blocks. We detail the characteristics of the execute-order-validate architecture below:

\begin{enumerate}[leftmargin=\parindent,align=left,labelwidth=\parindent,labelsep=2pt, topsep=0pt, partopsep=0pt]

\item \textit{Execution Phase}: A client sends a cryptographically signed transaction proposal to the peers. Upon receiving the transaction proposal, peers \textit{simulate} the transaction, i.e., they run the required chaincode on their own copy of the ledger without updating their ledger or sending a corresponding update to other peers. After the simulation, peers respond to the client with an \emph{endorsement}. The endorsement includes the peer's digital certificate chain, confirming their organization membership, and the peer's digital signature on the transaction proposal, the transaction's read-write set at the time of simulation (which details the version numbers but not the values of the variables the transaction interacts with~\cite{fabric-rwset}), and the outcome of the simulation.

\item \textit{Ordering Phase}: Once a client has collected sufficient endorsements for a proposal to be deemed acceptable by honest peers, it packs these endorsements into a transaction and forwards it to the ordering service. The orderers use a consensus protocol, such as Kafka~\cite{kreps2011kafka} or \ac{raft}~\cite{184040}, to order the transactions, grouping them into batches (blocks), and signing them without evaluating their validity. Upon reaching consensus on a block, they broadcast it to a subset of peers (one \emph{anchor peer} for each organization) for validation.

\item \textit{Validation Phase}: Once a peer receives a block -- either directly from an orderer or via a gossip protocol from a fellow peer belonging to the same organization -- it validates the content in three steps~\cite{foschini2020hyperledger}:

    \begin{enumerate} [leftmargin=\parindent,align=left,labelwidth=\parindent,labelsep=2pt, topsep=0pt, partopsep=0pt]

    \item First, every transaction within a block is subjected to parallel verification through the \ac{vscc}, which ensures that they have accumulated the appropriate number of endorsements as dictated by the endorsement policy. The \ac{vscc} also verifies that all executions came to the same result. If a transaction fails any of these tests, \ac{vscc} marks it as invalid. 

    \item Next, valid transactions undergo \ac{mvcc} -- a sequential check of whether the simulations were conducted on compatible versions of the ledger. Should any variable within the transaction's read-set have been altered in the peer's ledger since its endorsement, the transaction is marked as invalid.
    
    \item Once a transaction in a block has passed the \ac{vscc} and \ac{mvcc} checks, the peer updates the corresponding variables in its state database in what is called the \emph{commit} step. Invalid transactions, while not influencing the ledger's state, are nevertheless recorded in the ledger.
    \end{enumerate}
    
\end{enumerate}

\subsection{Quorum}
\label{quorum}

Quorum is another blockchain that has emerged as a significant player among the industry's permissioned blockchains~\cite{alastria,vakt}.  Quorum is based on Geth, an Ethereum client, and, as a result, leverages many of its characteristics. In contrast to Fabric, it utilizes a single, deterministic smart contract language, Solidity. This choice streamlines development, as Solidity is the primary language used in the broader, permissionless blockchain space \cite{defillama}, making it simpler to transfer applications from the public permissionless Ethereum blockchain to Quorum. Quorum relies on the more common \textit{order-execute} architecture under which transactions are first ordered and batched into blocks using a consensus mechanism and only subsequently executed.

Quorum supports four different consensus mechanisms: \ac{raft}, \ac{qbft}, \ac{ibft}, and Clique~\cite{quorum_readme}.
The details around the architecture vary slightly depending on the consensus mechanism used. Under \ac{raft} (the consensus mechanism used later in our analysis), each node can take only one of two roles: leader~(minter) or follower~(verifier). There is only one leader in each network.
Below, we detail the characteristics of the order-execute architecture based on the \ac{raft} consensus mechanism:

\begin{enumerate}[leftmargin=\parindent,align=left,labelwidth=\parindent,labelsep=2pt, topsep=0pt, partopsep=0pt]

\item \textit{Ordering Phase}: Clients send cryptographically signed transactions to the nodes. Subsequently, the nodes perform preliminary validations on the transactions, such as verification of the transactions' attributes (e.g., nonce, syntax, signatures, etc.). Once verified, the nodes disseminate the pre-validated transactions to other nodes via a gossip protocol, allowing them to be added to each node's transaction pool (\emph{mempool}). Once the transaction reaches the leader, it orders valid transactions according to the network's established priority rules. Finally, the ordered transactions are compiled into a block, which is disseminated across the network to follower nodes.

\item \textit{Execution Phase}: Upon receiving the block, followers attach the block to their version of the blockchain and send a message to the leader to confirm acceptance of the block. After receiving acceptance messages from the majority of nodes, the block is considered valid and becomes the new head of the blockchain.
\end{enumerate}

\subsection{Distributed Ledger Performance Scan}
\label{dlps}

The \ac{dlps} is designed as an end-to-end pipeline in which users can define the benchmarking specifications both for the configuration of the blockchain and client network, as well as the experiment settings~\cite{sedlmeir_dlps_2021}. All the necessary parameters can be set using a single config file, and we simplified further the deployment of \ac{dlps} across different computing environments by dockerizing its launch on local machines. Additionally, we extended its graphical capabilities to allow for a more in-depth analysis. The \ac{dlps} framework is composed of three Python packages, BlockchainFormation, DAppFormation, and ChainLab, which automatically set up the blockchain networks' nodes, smart contracts, and clients and iteratively execute measurements and collect and analyze data based on which they steer the benchmarking process. The benchmarking follows a recursive localization of maximum throughput with growing resolution \citep[for more details, see][]{sedlmeir_dlps_2021}.

This process is implemented by gradually increasing the request rate directed at the blockchain to push the network to its throughput limit~\cite{sedlmeir_dlps_2021}. The first localization run starts by sending asynchronous requests at a specified base rate. The \ac{dlps} measures the time at which clients send their transaction requests and receive successful notifications of their execution and determines average \ac{freq} and \ac{fresp} across all clients. If the network manages to keep up with \ac{freq} for a certain duration, this test concludes, and the next one starts with an increased request rate. If \ac{fresp} falls behind~\ac{freq} by more than a predetermined threshold (e.g., 5\,\%), a specified number of retries is performed. Should all retries fail, the ramping-up sequence is terminated, and the next ramping-up sequence for localization commences, starting from a base rate slightly lower than the maximum throughput achieved previously (e.g., 80\,\%). Further rampings can follow with smaller increments. This increases the granularity of measurements and data collection around the rate at which the system failed initially, allowing for the determination of a network's behavior close to maximum sustainable throughput. Once the experiment is complete, the \ac{dlps} generates multiple figures summarizing the gathered information. The \ac{dlps} also stores the benchmarking data in CSV files, allowing users to perform their own analyses. We make heavy use of these CSV files for the subsequent analysis.

\section{Related work}
\label{sec:related}

Evaluating the performance and assessing the capabilities of permissioned blockchains is vital~\citep{guggenberger_-depth_2022}. Yet, there is only little attention on this subject, and the available material tends to be highly fragmented and heterogeneous, making the comparison and reproducibility challenging~\citep{geyer2023end}. For instance, in the case of Fabric, publications investigate different versions, different consensus mechanisms, and heterogeneous network configurations and hardware. 
The use of different benchmarking tools aggravates these inconsistencies~\cite{javaid_optimizing_2019}, making it very difficult for practitioners to utilize the results for their blockchain projects. The majority of research also focuses on evaluating overall blockchain performance characteristics and puts less emphasis on identifying possible bottlenecks~\cite{guggenberger_-depth_2022, baliga_performance_h_2018}.

\begin{figure}[!b]
    \centering
    \includegraphics[width=\linewidth, trim=7.1cm 2cm 4.75cm 0.5cm, clip]{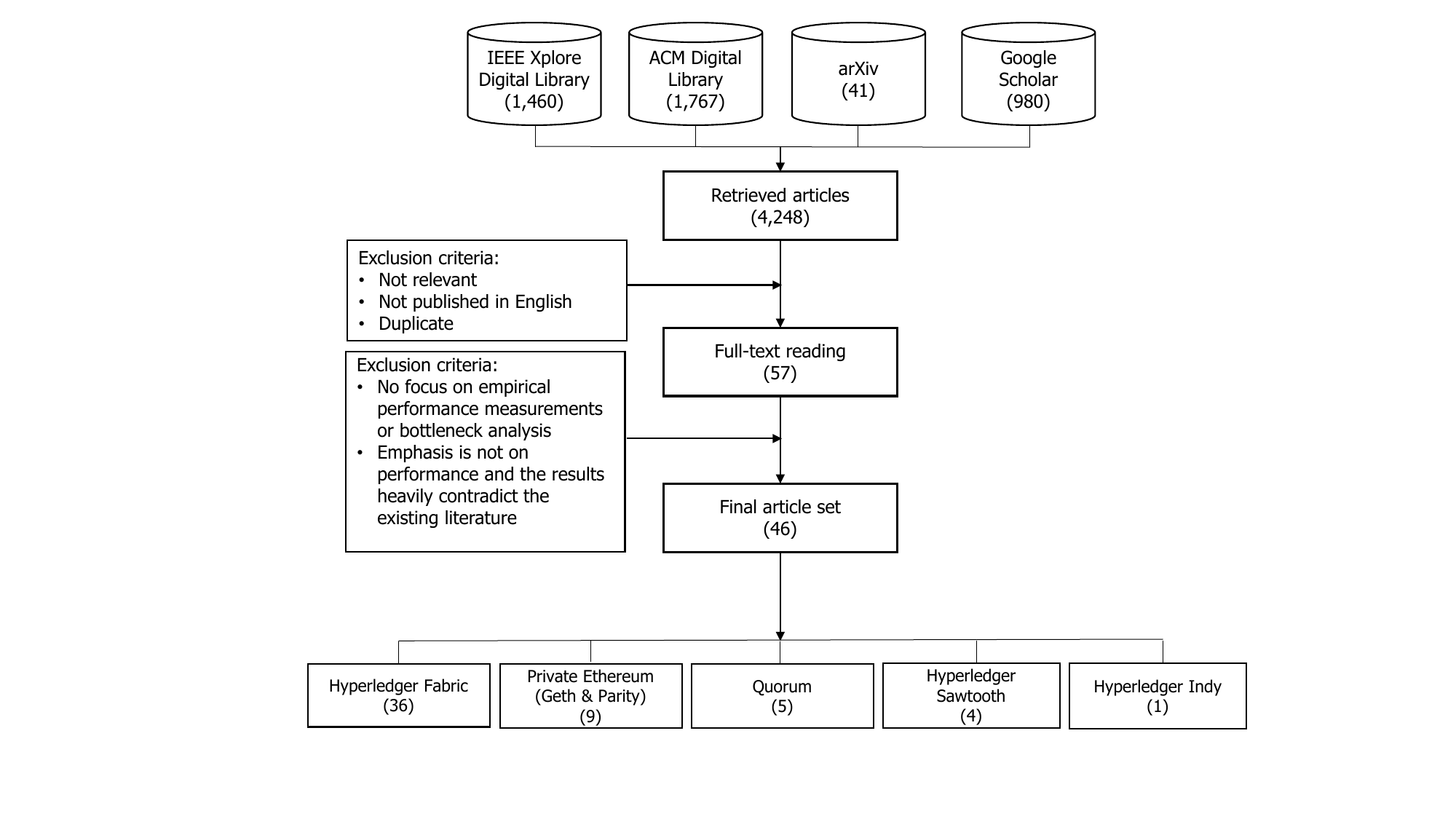}
    \caption{Overview of our systematic literature review.}
    \label{fig:litrev}
\end{figure}

To gain a full overview of the benchmarking academic literature, we started by performing a systematic literature review to identify publications that evaluate blockchain performance in general and the subset that provides insights into bottlenecks through in-depth analyses. We used the broad search string \emph{(blockchain OR ``distributed ledger technology'') AND (performance OR throughput OR latency) AND (benchmarking OR measurement OR evaluation OR analysis)} on Google~Scholar, ACM~Digital~Library, IEEE~Xplore, and arXiv. Our search yielded 4,248 results. Google Scholar alone provided approximately 57,000 results, but due to the limitations of the platform~\cite{gusenbauer2019google}, we could only access 980 of those. After filtering the results through a manual review of titles and abstracts and removing duplicates, 57~publications remained. For these, we performed a full-text screening and excluded the articles that do not provide empirical insights on blockchain performance, for example, because they focus on performance modeling, and articles in which performance evaluation is only a supplement and where the results contradict these in publications with a heavy benchmarking focus. Finally, we ended up with 46~relevant publications relating to five different blockchains. We found 36 publications that include measurements on Fabric~\cite{xu_latency_2021, wang_performance_2020, 10.1007/978-3-031-16092-9_1, thakkar_performance_2018,thakkar_scaling_2021, sharma_how_2018, shalaby_performance_2020, sedlmeir_dlps_2021, samy_enhancing_2021, pongnumkul_performance_2017, 8845168, nasirifard_fabriccrdt_2019, nasir_performance_2018, nakaike_hyperledger_2020, monrat_performance_2020, liu_adaptive_2021, kuzlu_performance_2019,klenik_porting_2021,javaid_optimizing_2019, hao_performance_2018, guggenberger_-depth_2022, gorenflo2020fastfabric, geyer_performance_2019, foschini2020hyperledger, dreyer_performance_2020, 10.1145/3318464.3389693, dinh_untangling_2018,dabbagh_performance_2020, 10.1145/3448016.3452823, bergman_permissioned_2020,baliga_performance_h_2018, androulaki_endorsement_2019, androulaki_hyperledger_2018, geneiatakis_blockchain_2020, joshi_blockchain_2019, dinh_blockbench_2017,sukhwani2018modeling}, 9~on private Ethereum (Geth and Parity)~\cite{dinh_blockbench_2017, leal_performance_2020, monrat_performance_2020, pongnumkul_performance_2017, rouhani_performance_2017, sedlmeir_dlps_2021, toyoda_function-level_2020, di_ciccio_performance_2019, benahmed_comparative_2019}, 5~on Quorum~\cite{baliga_performance_q_2018,mazzoni_performance_2021, monrat_performance_2020, sedlmeir_dlps_2021, shapiro_performance_2020}, 4~on Hyperledger Sawtooth~\cite{benahmed_comparative_2019,sedlmeir_dlps_2021, moschou_performance_2020, shi_operating_2019} and one on Hyperledger Indy~\cite{sedlmeir_dlps_2021}. Figure~\ref{fig:litrev} features an overview of our literature research.

In the case of Fabric, we do not take into consideration research performed on Fabric~v0.6~since it used the order-execute architecture. For the subsequent versions, several papers provide key insights on potential bottlenecks. For Fabric~v~1.0 and in one of the first detailed benchmarking attempts, Androulaki~et~al.~\cite{androulaki_hyperledger_2018} identify the validation phase and, in particular, the \ac{vscc} as a major bottleneck. Thakkar~et al.~\cite{thakkar_performance_2018}, using different Fabric configurations, find three major bottlenecks of~v1.0, all of which relate to the validation step: the process of verifying certificates in the \ac{vscc} to ensure that the endorsement policy is fulfilled, the sequential validation of transactions (\ac{mvcc}), and the multiple calls to the StateDB (when using CouchDB) during the validation and commit steps. Ruan~et~al.~\cite{10.1145/3318464.3389693} using~v1.3, again point toward the validation phase without specifying which component, as the bottleneck, especially when many unserializable transactions are included in the ledger, leading to higher latency and slowing down validation.

Wang~et~al.~\cite{wang_performance_2020} find that for v1.4, the \ac{vscc} execution during the validation phase remains the bottleneck as parallelization is limited and, consequently, throughput does not scale well with the number of cores. Chacko~et~al.~\cite{10.1145/3448016.3452823} identify that transaction failures point towards two systemic issues independent of configuration, with the validation stage acting as a likely bottleneck for~v1.4. Specifically, they find that \ac{mvcc} read conflicts, which occur when changes in the world state impact a transaction after it has already gained the required endorsements, result in transaction failure, necessitating a return to the execution phase for a new round of endorsements. The second issue lies with endorsement policy failures caused by inconsistencies between ledger copies across different peers, significantly slowing down the \ac{vscc} and, as a result, transaction processing. For the same Fabric version, Gorenflo~et~al.~\cite{gorenflo2020fastfabric} and Thakkar~et~al.~\cite{thakkar_scaling_2021} provide evidence that significant changes to the validation process could considerably improve throughput, further supporting the notion that the bottleneck lies within the validation phase.  In summary, across all examined versions of Fabric, the consensus among researchers is that the main bottleneck lies within the validation stage. It is also worth noting that we could not find research focusing on identifying bottlenecks in Fabric~v2.0 or higher.

When it comes to Quorum, Mazzoni~et~al. \cite{mazzoni_performance_2021} posit that a potential bottleneck lies with the node's \ac{rpc} server buffers being capped at~128KB.  While this limitation suggests a maximum transaction size of 128KB, it is improbable that this limit constitutes the main bottleneck under normal circumstances, as the average transaction sizes usually amount to only a few hundred bytes. For Geth, Toyoda~et~al.~\cite{toyoda_function-level_2020} identify that the primary bottleneck stems from the poor utilization of the nodes' multi-threading and calls of \texttt{crypto.Ecrecover} -- a function for retrieving public keys from signatures that is called every time a transaction reaches a node. We found no study on private Ethereum with Aura consensus (Parity), Hyperledger Sawtooth, and Hyperledger Indy that aimed to determine their performance bottlenecks.

\section{Method and Results}
\label{sec:m&f}

We developed the overall approach by analyzing the experiments described in related work, using the~\ac{dlps} to collect data and~\ac{eda} to identify bottlenecks. We collected a large volume of data on factors influencing node performance, ensuring a comprehensive overview of their resource utilization. After cleaning and validating the data, we used~\ac{eda} in search of irregularities or drops in the blockchain’s performance. Upon identifying these issues, we conducted a detailed examination of the relevant metrics at a finer granularity to determine the root causes of these behaviors.
We divided our analysis into two major components. The first part identifies potential bottleneck candidates by examining the different node resources that can impact their performance. In the second part, we examine the relationship between these candidates and the blockchain’s throughput in varying degrees of resolution, for instance, narrowing down the time window or selection of components.
So far, most enterprise applications seem to have focused on Fabric and Quorum, and related work indicates that the bottleneck analysis of both blockchains is intricate, so we decided to detail our bottleneck identification method for both blockchains. 

For our analysis, we selected a particular experiment for Fabric v2.0, drawing upon the findings of~\cite{guggenberger_-depth_2022} to determine a configuration that seems robust under modifications and extended it to Quorum~v23.4. For Fabric, the network configuration comprised 16~clients, 8~peers, and 4~orderers, with four organizations comprised of two peers each. In contrast, the Quorum configuration consisted of 16~clients and 8~nodes. For both blockchains, we opted for the \ac{raft} consensus mechanism due to its minimal overhead, as previous publications suggest that consensus is not the bottleneck in this case~\citep{guggenberger_-depth_2022, mazzoni_performance_2021}. We conducted the experiment on the AWS~cloud platform (Amazon~EC2), where each node was configured in an independent EC2~instance allocated with 16~vCPUs, 64\,GB~of~RAM, and 1\,Gbps of bandwidth running Ubuntu Server~18.04~LTS~(HVM).

\subsection{Fabric: Resource utilization}
\label{resource_util}

\begin{figure}
\centering
\includegraphics[width=1\linewidth]{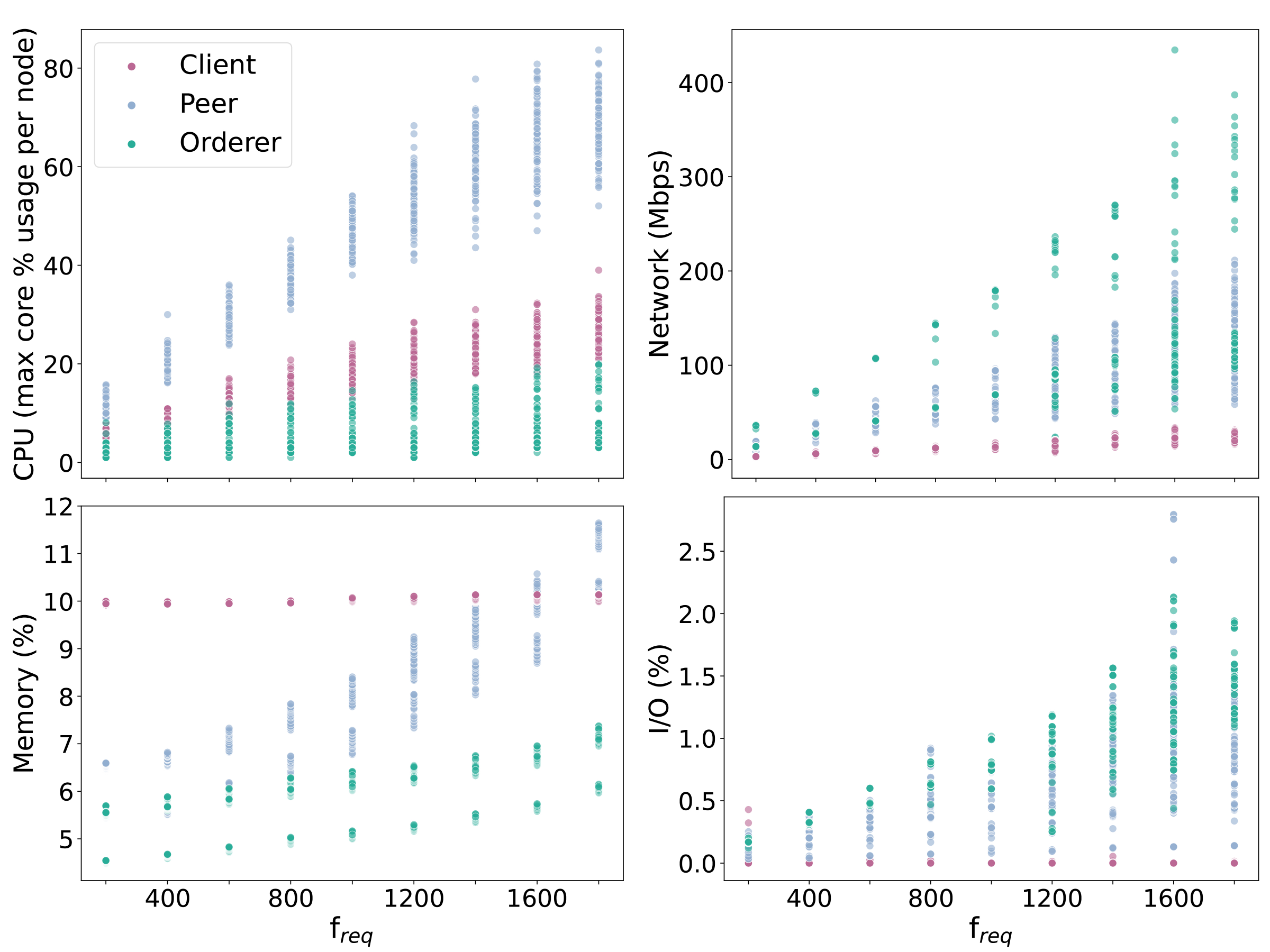}
\caption{Fabric -- Key resource utilization for different request rates.}
\label{fig:Scatter_resources}
\end{figure}

The initial analysis focuses on understanding the impact of incremental increases in the request rate on the resource utilization of each node. This is achieved by plotting a data point for each node every second during a 14-second time-frame within a 20-second experiment (excluding the initial and final 3~seconds where the ramp-up and ramp-down of \ac{freq} may cause temporary effects), corresponding to different request frequencies (Figure~\ref{fig:Scatter_resources}). This method aims to uncover potential trends and correlations between resource usage and increased \ac{freq}. This initial step already indicates that peers exhibit the highest strain in \ac{CPU} usage, as demonstrated by the maximum utilization percentage across all cores on each node while ordering and client nodes display significantly lower \ac{CPU} utilization rates.

In terms of memory usage, both clients and ordering nodes appear largely unaffected by increases in the request rate. However, peers show signs of impact, albeit without a clear pattern. Regarding network and hardware utilization, client nodes maintain stable performance, whereas peer and orderer nodes demonstrate a moderate correlation between increased utilization and \ac{freq}, but as before, the exact trend is not clear. Note that at this early stage of the analysis, any anomalies could also suggest inaccuracies in the experimental setup. Therefore, it is crucial to eliminate this possibility before conducting a more detailed analysis to determine the structural causes behind these observations. Upon ruling out experimental errors, a separate analysis is conducted for each component.

\subsubsection{\textbf{CPU}}
\label{cpu}

The analysis begins by focusing on the average \ac{CPU} utilization across a 14-second period, with a subsequent focus on more detailed time intervals. We observe that as the request rate increases, the linear correlation between the request rate and CPU utilization of peers breaks at \ac{freq}=1600\,s$^{-1}$, indicating a potential saturation point or limitation in \ac{CPU} capacity. This poses a potential bottleneck that requires further examination. 

\begin{figure}
     \centering
     \makebox[\columnwidth]{\includegraphics[width=0.75\columnwidth, trim=0cm 0cm 0cm -0.5cm, clip]{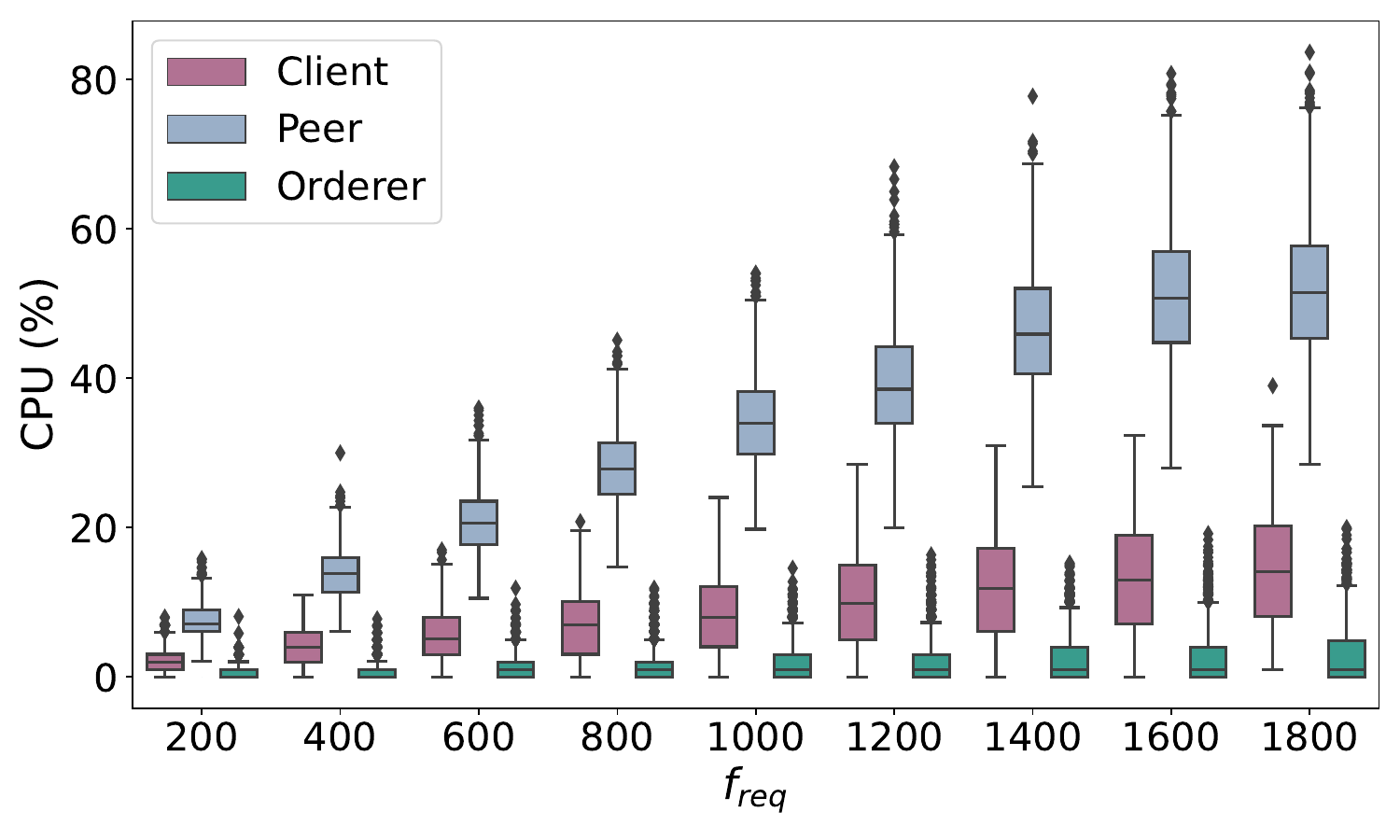}}
     \caption{Fabric -- CPU utilization of all cores.}
     \label{fig:Boxplot_all CPUs}
\end{figure}

Looking into the CPU usage across all individual cores (Figure~\ref{fig:Boxplot_all CPUs}) reveals limited utilization across orderers and clients that is not plateauing at higher request rates. Focusing on peers, we observe a significant variance in \ac{CPU} utilization among peers, ranging from 25\% to 80\% at higher request rates. Such variability could be an indicator that the network is distributing its resources unevenly, with some peers carrying more workload than others, or reflect an uneven allocation between nodes’ cores.

\begin{figure}
    \subfloat[Mean utilization per peer.\label{fig:CHART_workload distribution_peer_CPU_absolute_agg=mean}]{{\includegraphics[width=0.5\linewidth]{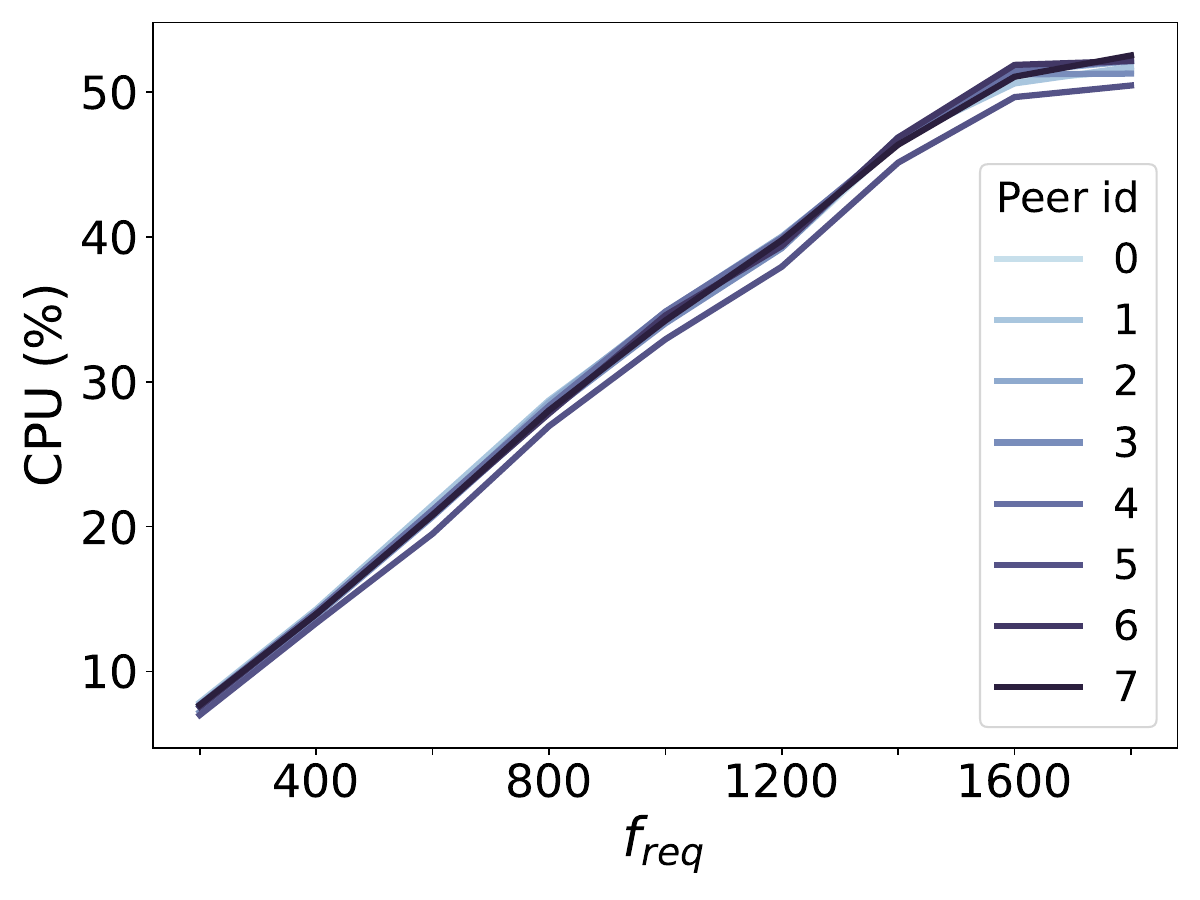}}}
    \subfloat[Mean utilization of peer~0 per core.\label{fig:CHART_workload distribution_peer0 processors}]{{\includegraphics[width=0.5\linewidth]{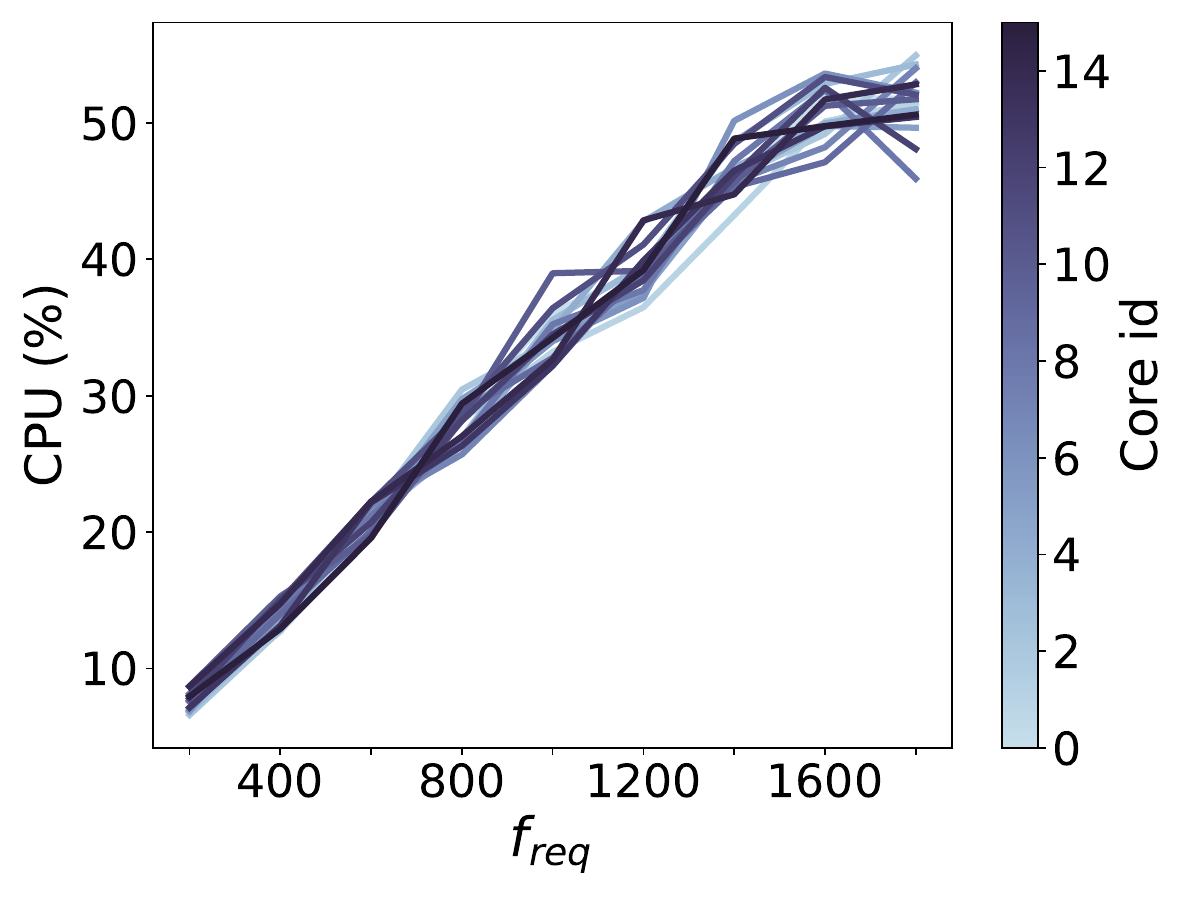}}}
    \caption{Fabric -- peer mean CPU utilization.}
    \label{fig:CPU utilization in peers}
\end{figure}

First, we investigate the CPU utilization per peer in Figure~\ref{fig:CHART_workload distribution_peer_CPU_absolute_agg=mean}, which suggests an equitable distribution of computational resources among the peers, with all of them showing comparable levels of computational effort, with peer~5 falling behind slightly. Subsequently, the investigation shifts to analyzing the mean \ac{CPU} utilization of individual cores for a single peer (peer~0 in this case) in Figure~\ref{fig:CHART_workload distribution_peer0 processors}. It is apparent that this also is not the cause of the high fluctuations in \ac{CPU} usage since all cores indicate similar utilization.

\begin{figure}
    \centering
    \includegraphics[width=0.6\linewidth, trim=0cm 0cm 0cm -1.8cm, clip]{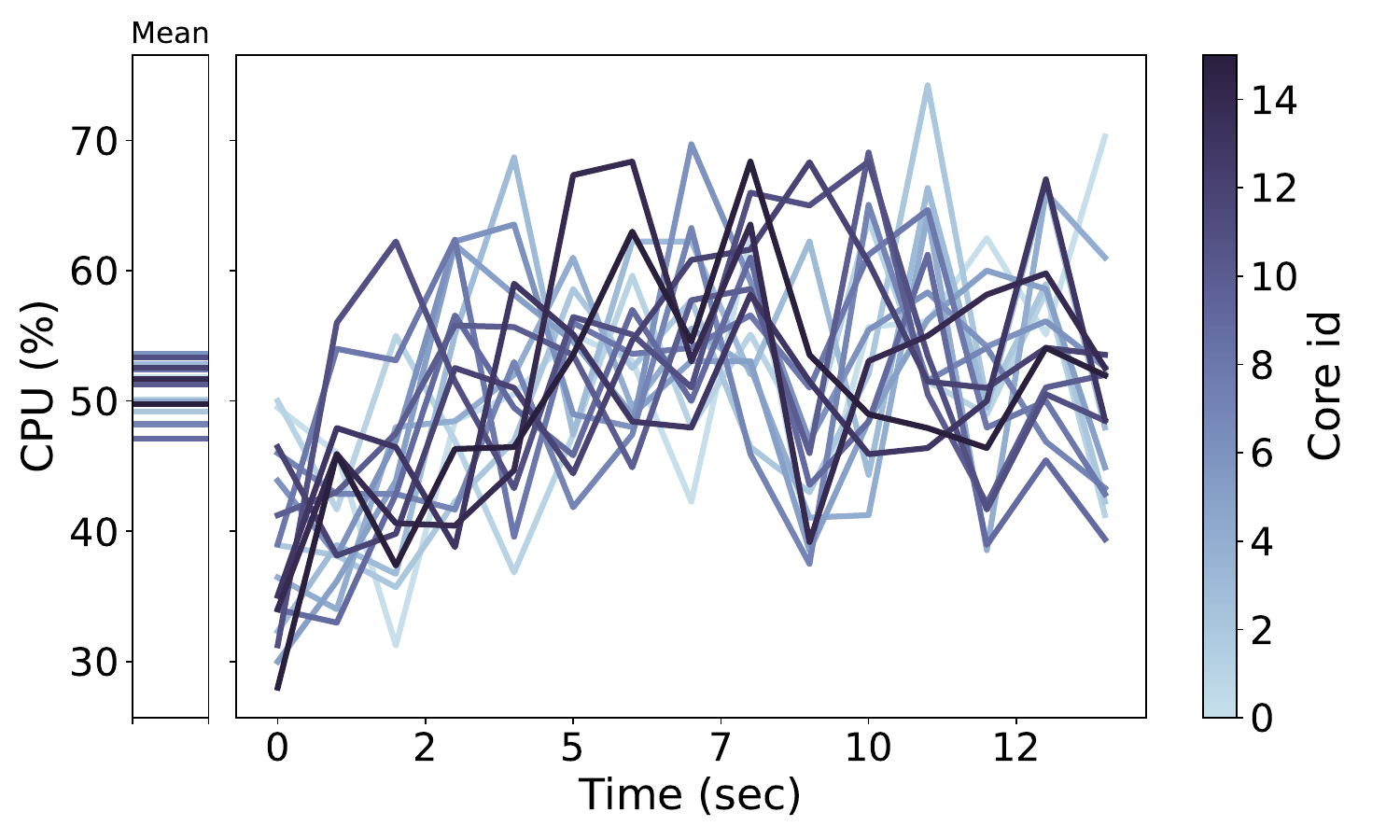}

    \caption{Fabric -- CPU utilization of peer~0 for \ac{freq}=1600\,s$^{-1}$.}
    \label{fig:Chart_timeline_peer0 processors freq=1600}
\end{figure}

Following the inconclusive results of both possible explanations, the analysis progresses to evaluate the temporal evolution of \ac{CPU} usage across individual cores at the highest \ac{freq} before the utilization plateaus (Figure~\ref{fig:Chart_timeline_peer0 processors freq=1600}). This \ac{freq} represents the peak stress on the network prior to any performance degradation. Here, we observe that individual cores can fluctuate as much as 30\,\% within a single run. At this point, this analysis comes to a halt as it is impossible to deduce the reasons behind these fluctuations from \ac{CPU} utilization data only. However, the mean utilization illustrates that these fluctuations balance themselves out over time, with all cores clocking in at a similar average. This suggests that the high fluctuations in \ac{CPU} usage are not the reason behind the plateauing.

It is also worth noting that from Figure~\ref{fig:CPU utilization in peers}, we see that average \ac{CPU} usage plateaus at around 50\,\% across all cores on all peers. This marks a significant improvement to the prior versions of Fabric. However, it is important to highlight that in scenarios utilizing a higher number of vCPUs (16~in our case), Fabric~v2.0 still has a mediocre mean utilization for a resource that can limit the entire system, which leaves room for further improvement. 

\subsubsection{\textbf{Network}}
\label{network}

\begin{figure}
\subfloat[Mean utilization per orderer\\(inbound traffic of orderer~1,~2,~3 \\ overlaps with their outbound traffic).\label{fig:workload distribution_orderer_network_absolute}]{{\includegraphics[width=0.5\linewidth]{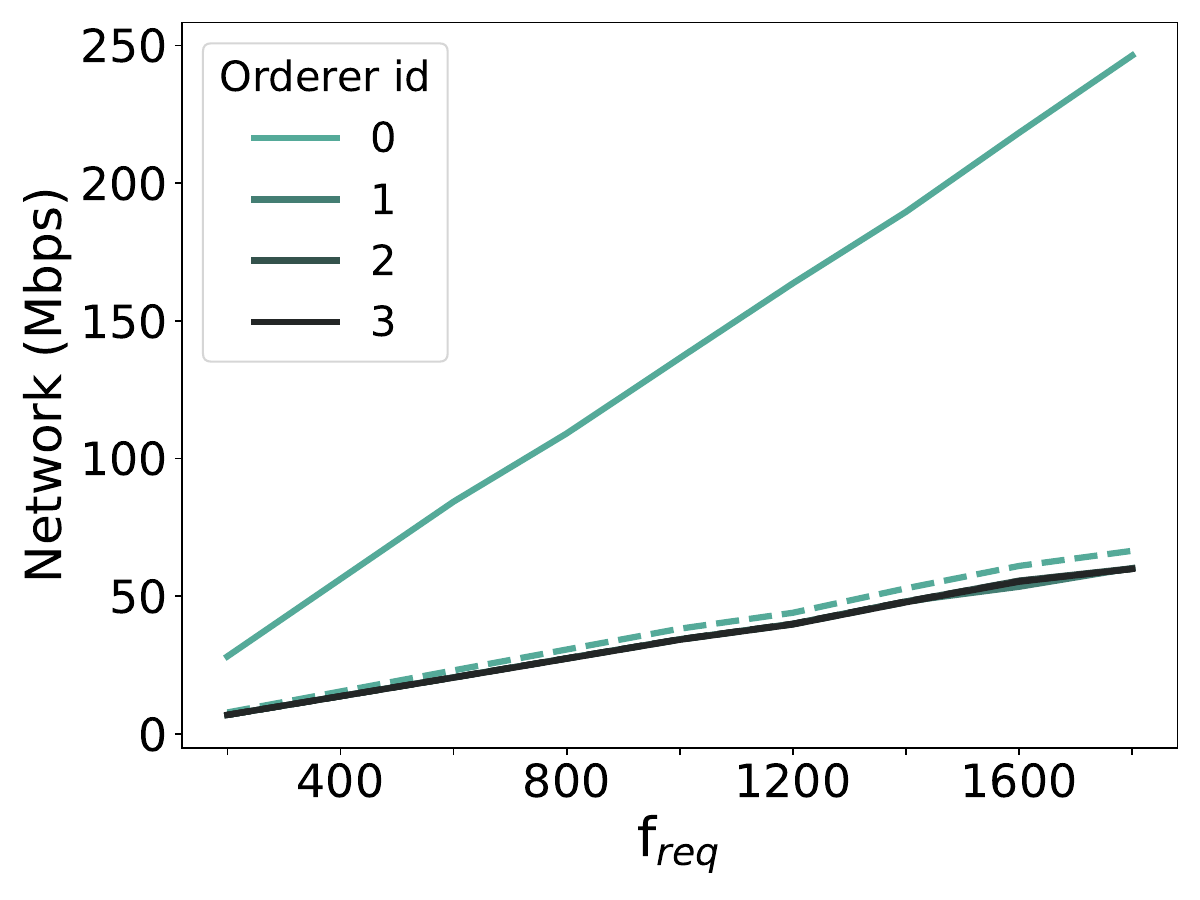} }}
\subfloat[Ratio of leader outbound traffic over follower outbound traffic.\label{fig:workload distribution_orderer_network out - leader durch follower}]{{\includegraphics[width=0.5\linewidth]{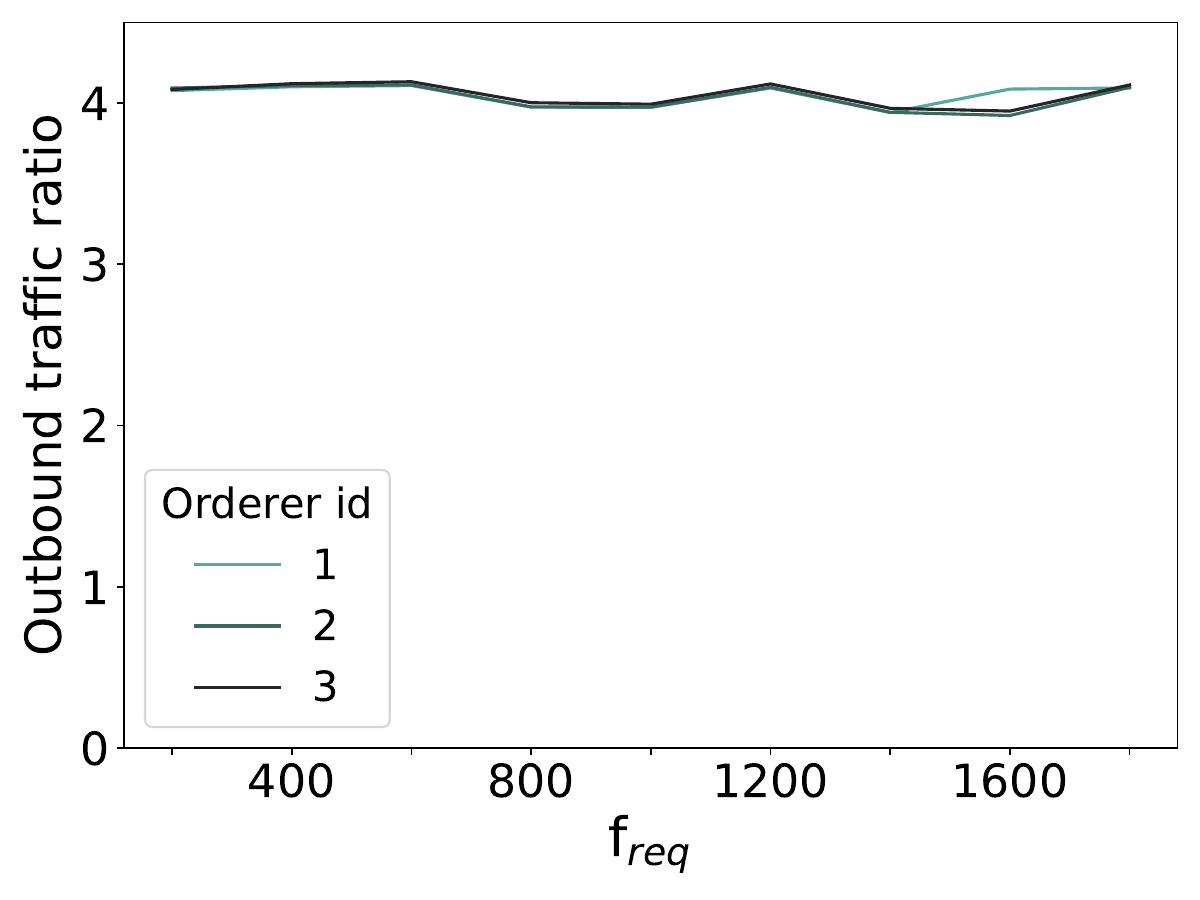} }}

\caption{Fabric -- orderer mean network utilization (outbound traffic as continuous and inbound traffic as dashed lines).}
\label{fig:network for orderers}
\end{figure}

The network-related part of the analysis begins by looking into the mean network utilization, distinguishing between inbound and outbound traffic for the different types of nodes. Unsurprisingly, there is a strong correlation between the request rate and mean traffic.
In the examination of orderers, we observe that both incoming and outgoing traffic do not plateau at high request rates, suggesting that they are not the bottleneck (Figure~\ref{fig:workload distribution_orderer_network_absolute}). Notably, orderer~0 broadcasts a disproportionate amount of traffic compared to the rest, which indicates that orderer~0 is the leader in \ac{raft} consensus and, as a result, has the additional task of broadcasting new blocks to each following orderer. In our experiments, we did not simulate crashing nodes, so orderer~0 remained the leader for the whole duration of the experiments (cf.~\cite{guggenberger_-depth_2022}), making this discrepancy logical.

A more detailed analysis of the network traffic allows its decomposition into individual components, such as the traffic generated by block propagation. According to Fabric's architecture, the outbound traffic of followers among orderers predominantly consists of sending blocks to the peers, and from Figure~\ref{fig:workload distribution_orderer_network_absolute}, an overlapping between inbound and outbound traffic of the followers is observed. The overlapping traffic, combined with the fact that followers receive each block once from the leader, suggests that each orderer forwards the received block only once.
To evaluate the validity of this hypothesis, we compare followers' outbound traffic with the leader's outbound traffic (Figure~\ref{fig:workload distribution_orderer_network out - leader durch follower}). The ratio consistently stands around four, aligning with our hypothesis and the Fabric architecture where the \ac{raft} leader distributes the block to each follower, accounting for three times in our case and an additional time to a peer, totaling four times. 
Our analysis does not take into consideration the traffic generated by consensus-related messages, such as appended entries or heartbeat messages~\cite{Fabric_config}, since differentiation between them and the traffic from block propagation is challenging. Nonetheless, the consensus-related messages generate significantly less traffic compared to block dissemination, and as a result, we expect that the outbound traffic of follower orderers provides a viable approximation for assessing the traffic associated with block propagation.

\begin{figure}
\subfloat[Mean utilization per peer.\label{fig:workload distribution_peer_network_absolute}]{{\includegraphics[width=0.5\linewidth]{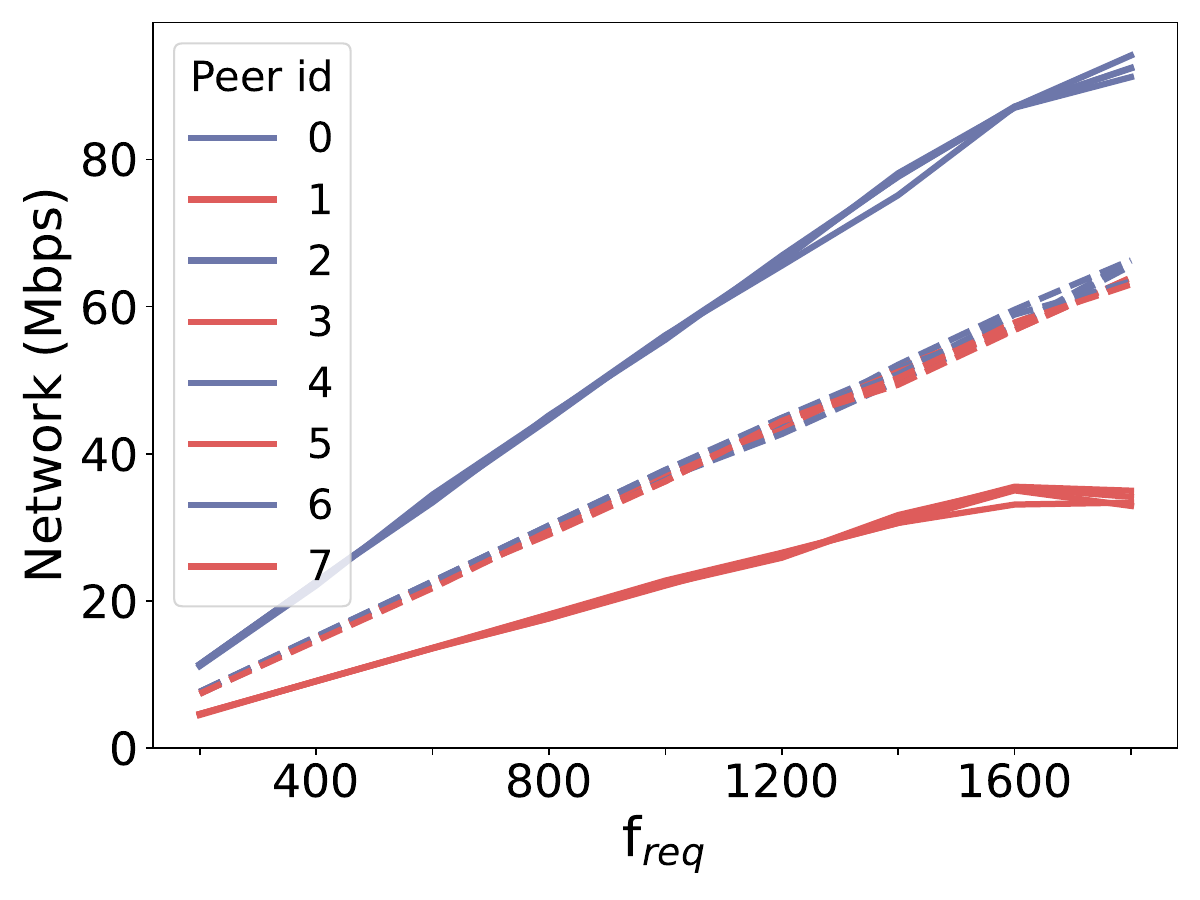} }}
\subfloat[Outbound traffic per peer excluding traffic generated by block propagation.\label{fig:plot_network_traffic_peer_no_block}]{{\includegraphics[width=0.5\linewidth]{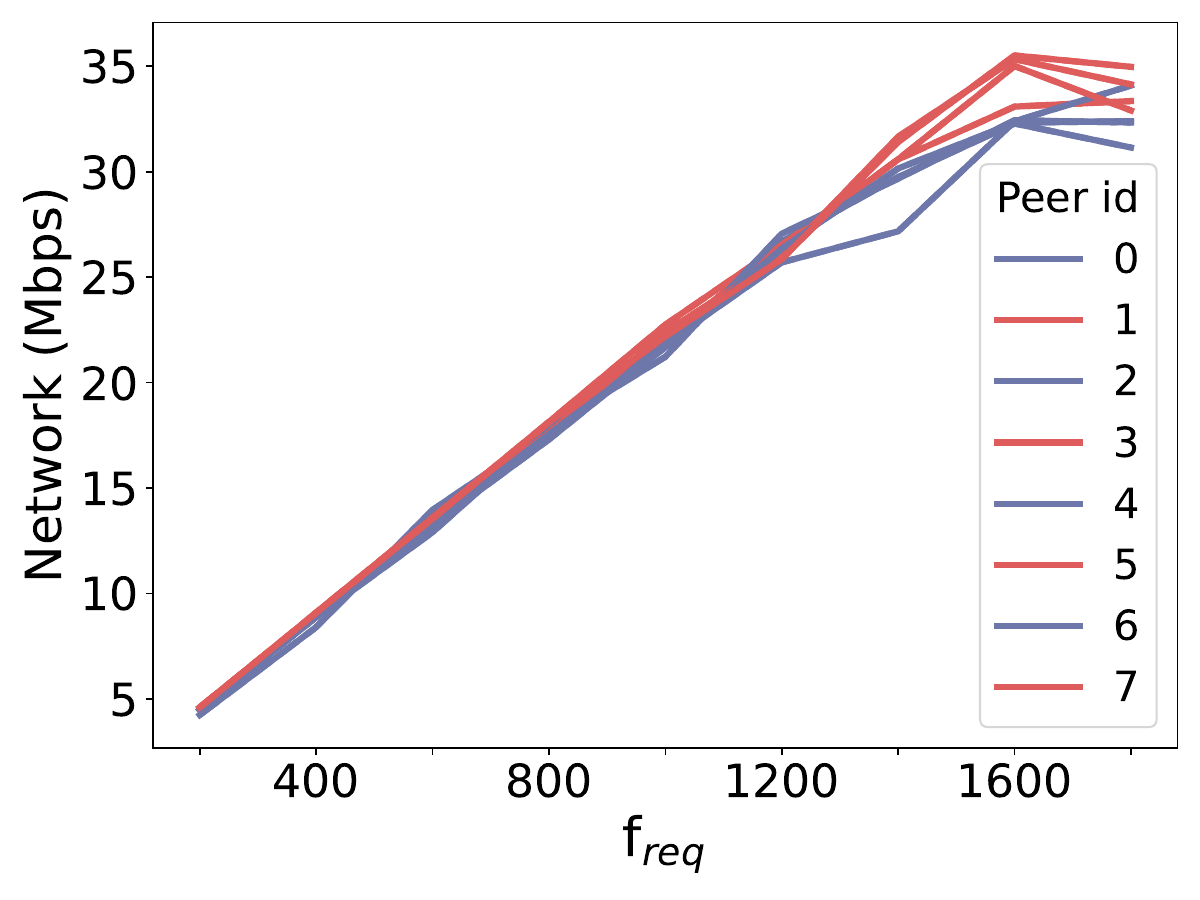} }}
\caption{Fabric -- peer mean network utilization (outbound traffic as continuous and inbound traffic as dashed lines).}
\label{fig:network for peers}
\end{figure}

Regarding peers (Figure~\ref{fig:network for peers}), we see that inbound traffic scales linearly with the \ac{freq} for all of them. This uniformity aligns with the architecture in our experiment, as all peers receive the same number of transaction proposals (i.e., endorsement requests) from clients and the same number of blocks. Additionally, inbound traffic scales linearly throughout the experiment, indicating it's not a bottleneck.
Concerning outbound traffic, we observe that it plateaus for some peers while remaining unaffected for others (Figure~\ref{fig:workload distribution_peer_network_absolute}). This divergence allows for the classification of peers into two main clusters, color-coded as blue and orange, based on their outbound traffic. The two groups distribute communication workload differently, with blue peers registering more than twice the sent data of the orange peers. The blockchain's architecture explains this divergence since blue peers are the gossip leaders of their respective organizations, with one gossip follower each. For instance, organization~0 consists of orderer~0, peer~0, and peer~1, with peer~0 acting as the anchor peer. As such, peer~0 receives all new blocks from orderer~0 and forwards them to peer~1.

The primary distinction in outbound traffic among peers stems from the block propagation between them. To confirm that blue peers correspond to gossip leaders and the accuracy of previous assumptions regarding traffic generated by block propagation, we perform a further analysis by deducting the traffic associated with block propagation, as obtained from the previous analysis of the ordering service, from the total outbound traffic of gossip leader peers (Figure~\ref{fig:plot_network_traffic_peer_no_block}).
The resulting traffic is relatively uniform across all peers, confirming that the blue peers are the gossip leaders. Furthermore, we observe a significant overlap in outbound traffic among peers, particularly at lower request frequencies, while some discrepancies exist at higher rates. This is expected as traffic stemming from consensus messages increases at higher request rates, making our approximation less accurate. The outbound traffic of peers consists of sending requested endorsements and confirmations that requested transactions have been committed to the blockchain to clients. Since the outbound traffic of peers plateaus at high \ac{freq}, it suggests that these components are potential bottlenecks.

To pinpoint the specific bottleneck, we focus on client traffic (Figure~\ref{fig:network for peers, orderers and clients}). The inbound traffic of clients, although it exhibits a drop at higher request rates, is generated by the same components as the outbound traffic of peers, leaving us with the same potential bottlenecks as before. The outbound traffic of clients is comprised of transaction proposals submitted to peers and endorsed transactions sent to the orderers. Since the traffic does not plateau, it suggests that none of them is the bottleneck.
Overall, we have seen that the traffic generated by the endorsed transactions the clients send to the ordering service and the traffic generated by the blocks the service sends to the peers never plateaus. This indicates that the execution and ordering phases are not the bottleneck. Since the endorsements that peers send back to the clients come in between the execution and ordering phases, this suggests that they are not the bottleneck, leaving us only with the transaction confirmations that peers send to clients as the possible culprit. These are sent after the validation phase, suggesting that this remains the primary bottleneck even in Fabric v2.0.

\begin{figure}
    \centering
    \includegraphics[width=0.725\linewidth]{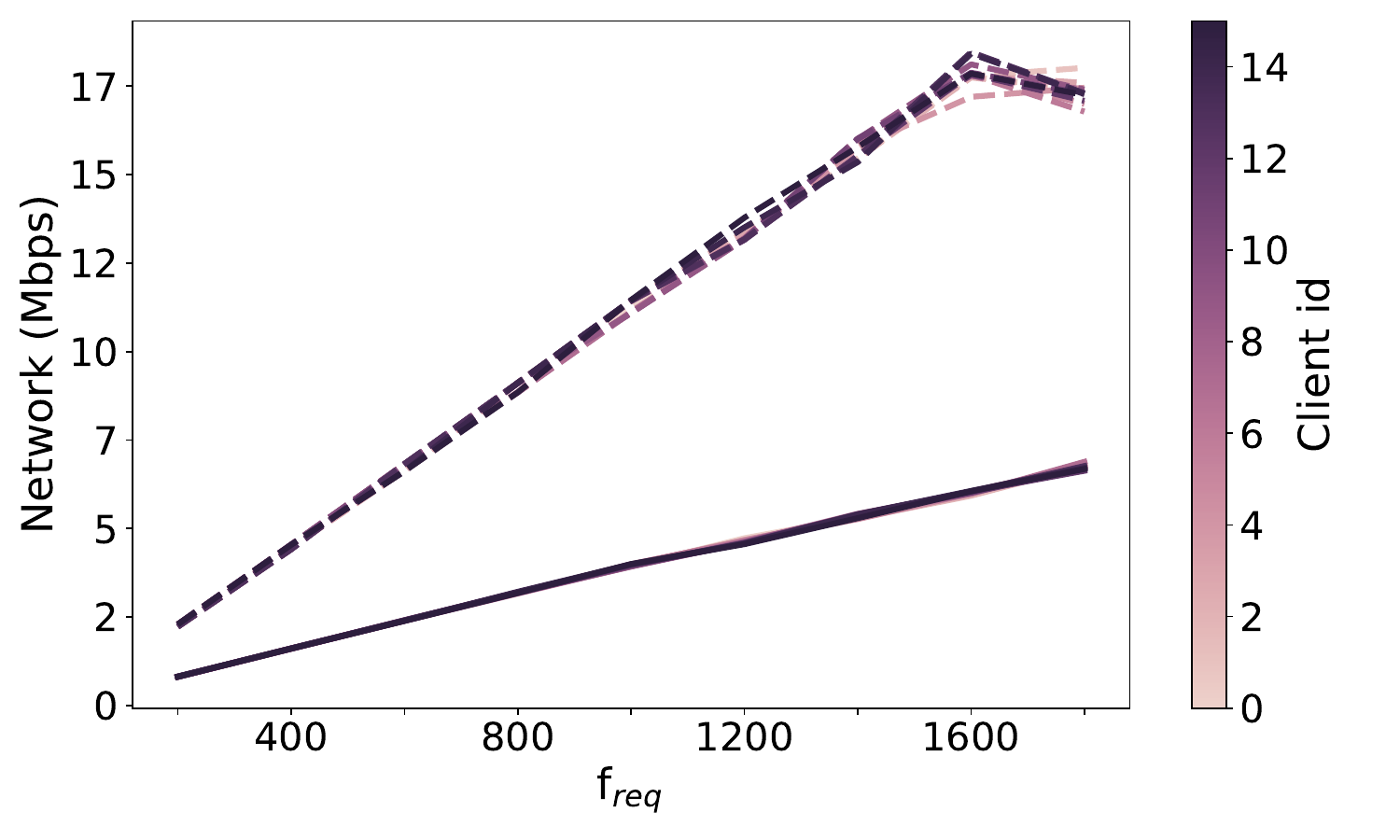} 

    \caption{Fabric -- Client mean network utilization (outbound traffic as continuous and inbound traffic as dashed lines).}
    \label{fig:network for peers, orderers and clients}
\end{figure}

\subsubsection{\textbf{Memory \& Hard Drive}}
\label{memory}

\begin{figure}
 \subfloat[Mean utilization per peer\label{fig:memory_usage_peer}]{{\includegraphics[width=0.5\linewidth]{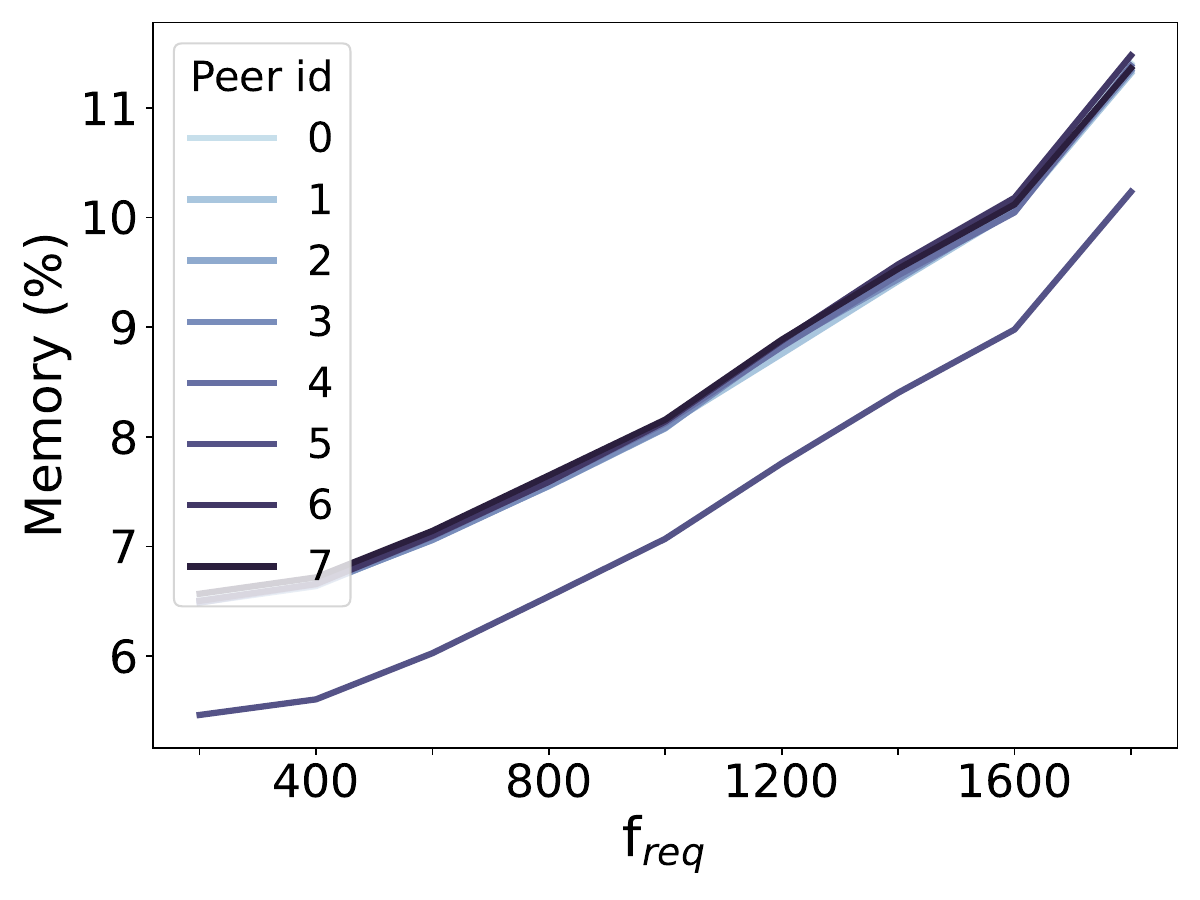} }}
 \subfloat[Mean utilization per orderer\label{fig:memory_usage_orderer}]{{\includegraphics[width=0.5\linewidth]{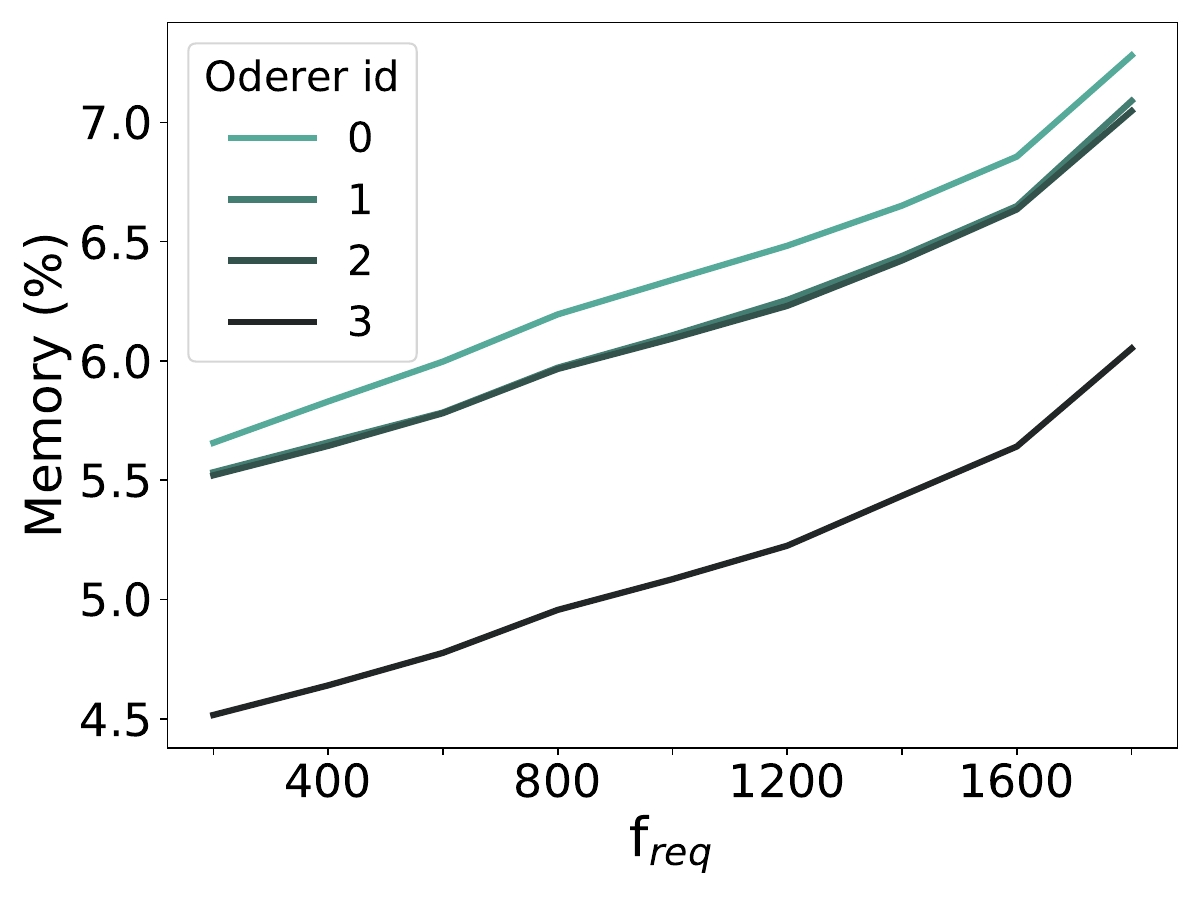} }}

\caption{Fabric -- Mean memory utilization.}
\label{fig:memory_utilisation}
\end{figure}

From Figure~\ref{fig:Scatter_resources}, we see that the utilization of memory and hard drive is limited, indicating that neither of them is the bottleneck. Despite this, we examine the resource utilization of peers and orderers further. We do not examine clients since their utilization remains unaffected by the increasing number of requests. 
Starting with memory usage (Figure~\ref{fig:memory_utilisation}), we see peers and orderers exhibit low and non-plateauing usage levels, suggesting that memory constraints are not the bottleneck. We observe some minor differences in memory utilization across orderers, but given the low overall utilization, it is highly unlikely that they constitute a bottleneck, and we don't investigate them further.

\begin{figure}
 \subfloat[Mean utilization per peer\label{fig:i_o_usage_peer}]{{\includegraphics[width=0.5\linewidth]{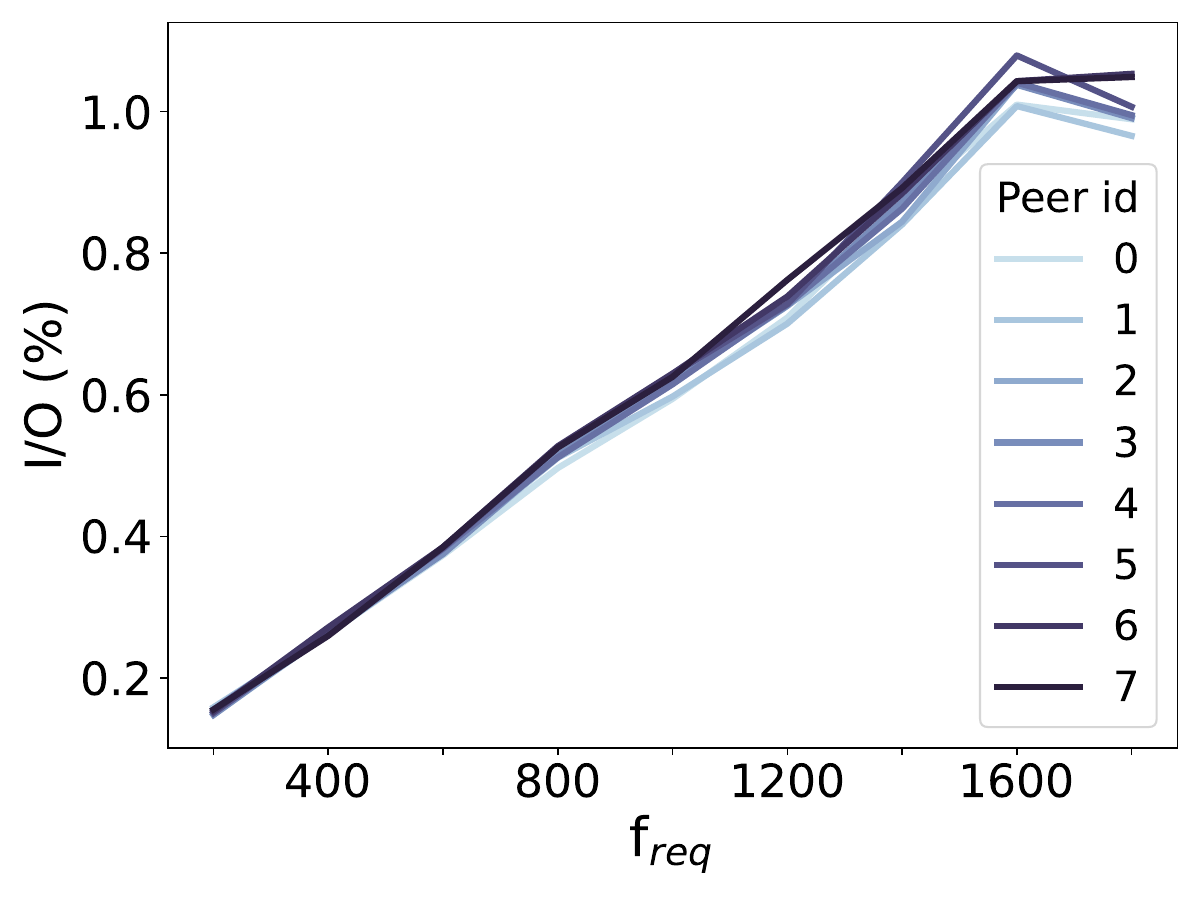} }}
 \subfloat[Mean utilization per orderer\label{fig:i_o_usage_orderer}]{{\includegraphics[width=0.5\linewidth]{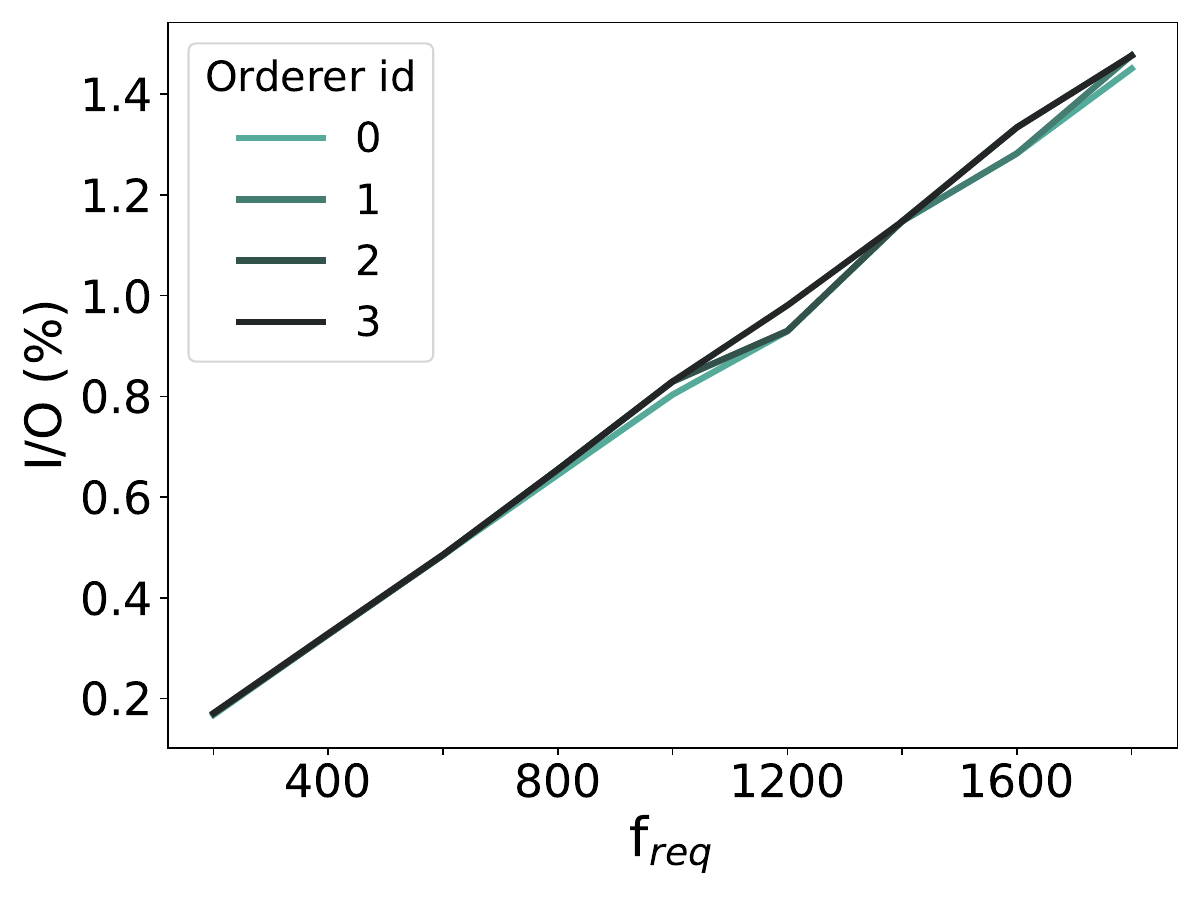} }}

\caption{Fabric -- Mean hard drive utilization.}
\label{fig:i_o_utilisation}
\end{figure}

Moving to hard drive utilization (Figure~\ref{fig:i_o_utilisation}), we observe that the orderers' usage is increasing linearly with the \ac{freq} and, as a result, likely does not pose a problem. Most peers' I/O operations (such as ledger updates or block processing) occur during or after the validation phase. As a result, even though their utilization plateaus, it does not provide us with any new information. Additionally, due to the variety in read/write operations that peers execute, it is impossible to differentiate between them with only data related to resource utilization at hand. Moreover, with the peak utilization at approximately 1~\%, it is evident that hard drive usage is far from reaching capacity, pointing out that constraints lie within a different component, which in turn limits hard drive utilization.

As mentioned in Section ~\ref{sec:background}, the validation phase is comprised of three steps: \ac{vscc}, \ac{mvcc}, and each peer updating its database. As the peers' hard drive utilization is minimal, this leaves only \ac{vscc} and \ac{mvcc} as the potential bottlenecks in the validation phase.

\subsection{Fabric: Throughput}
\label{throughput}

\begin{figure}
 \subfloat[1s window\label{fig:window_1s}]{{\includegraphics[width=0.5\linewidth]{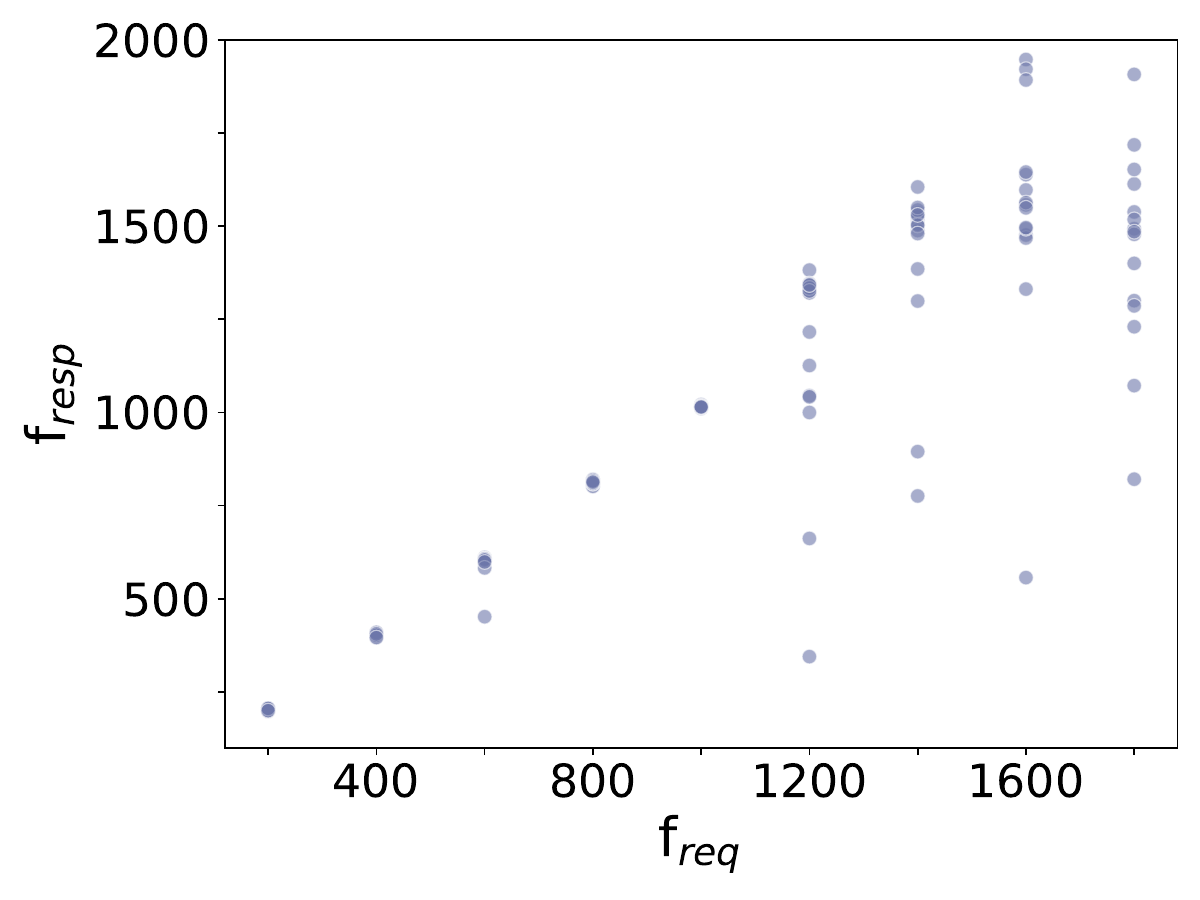} }}
 \subfloat[3s window\label{fig:window_3s}]{{\includegraphics[width=0.5\linewidth]{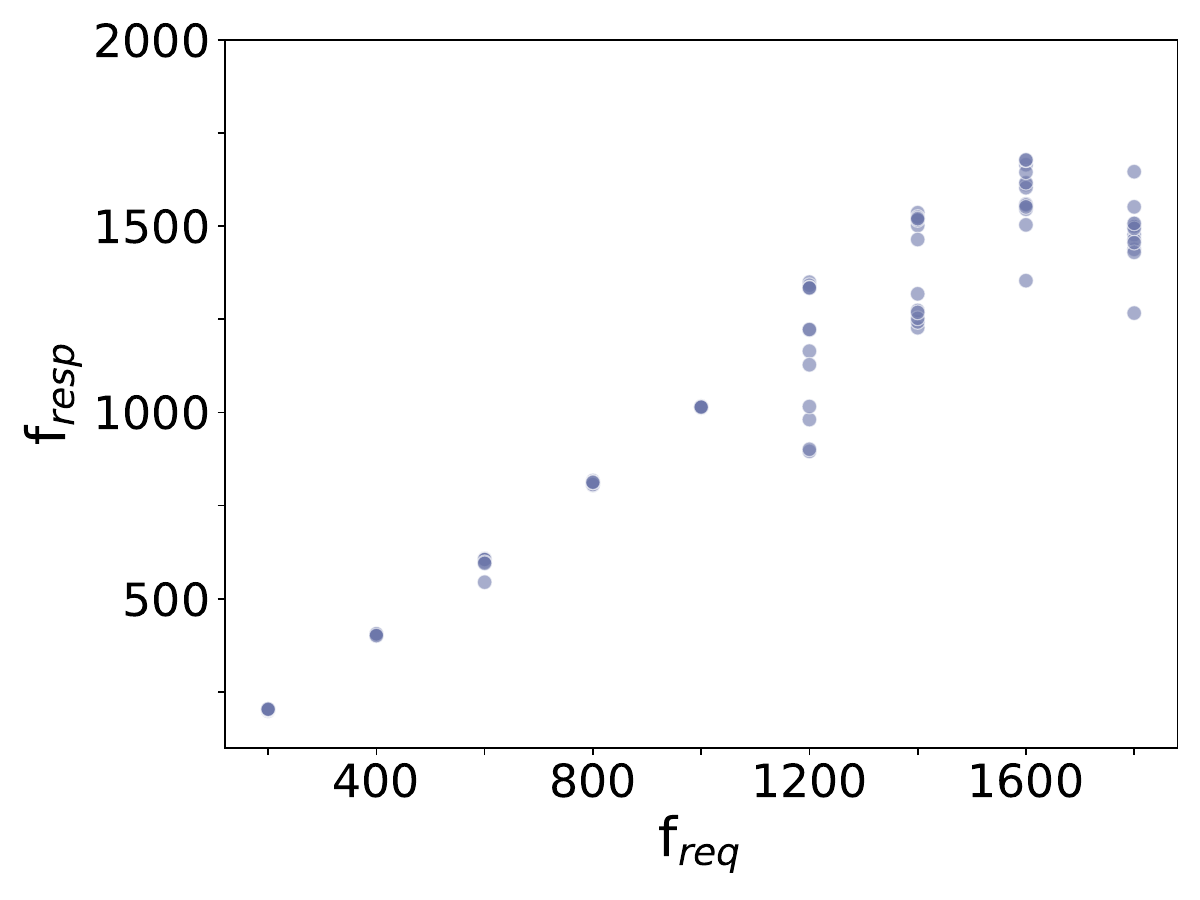}}}

  \subfloat[8s window\label{fig:wwindow_8s}]{{\includegraphics[width=0.5\linewidth]{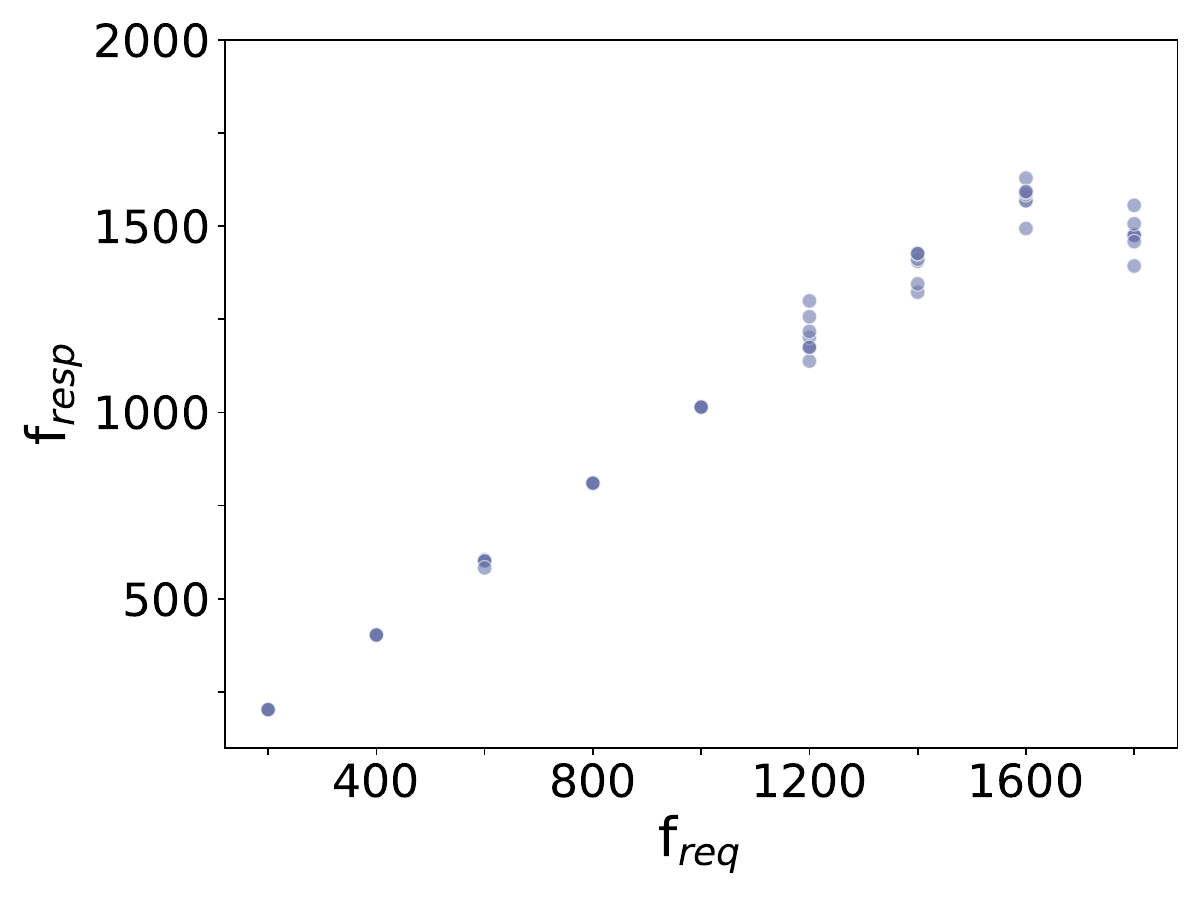} }}
 \subfloat[15s window\label{fig:window_14s}]{{\includegraphics[width=0.5\linewidth]{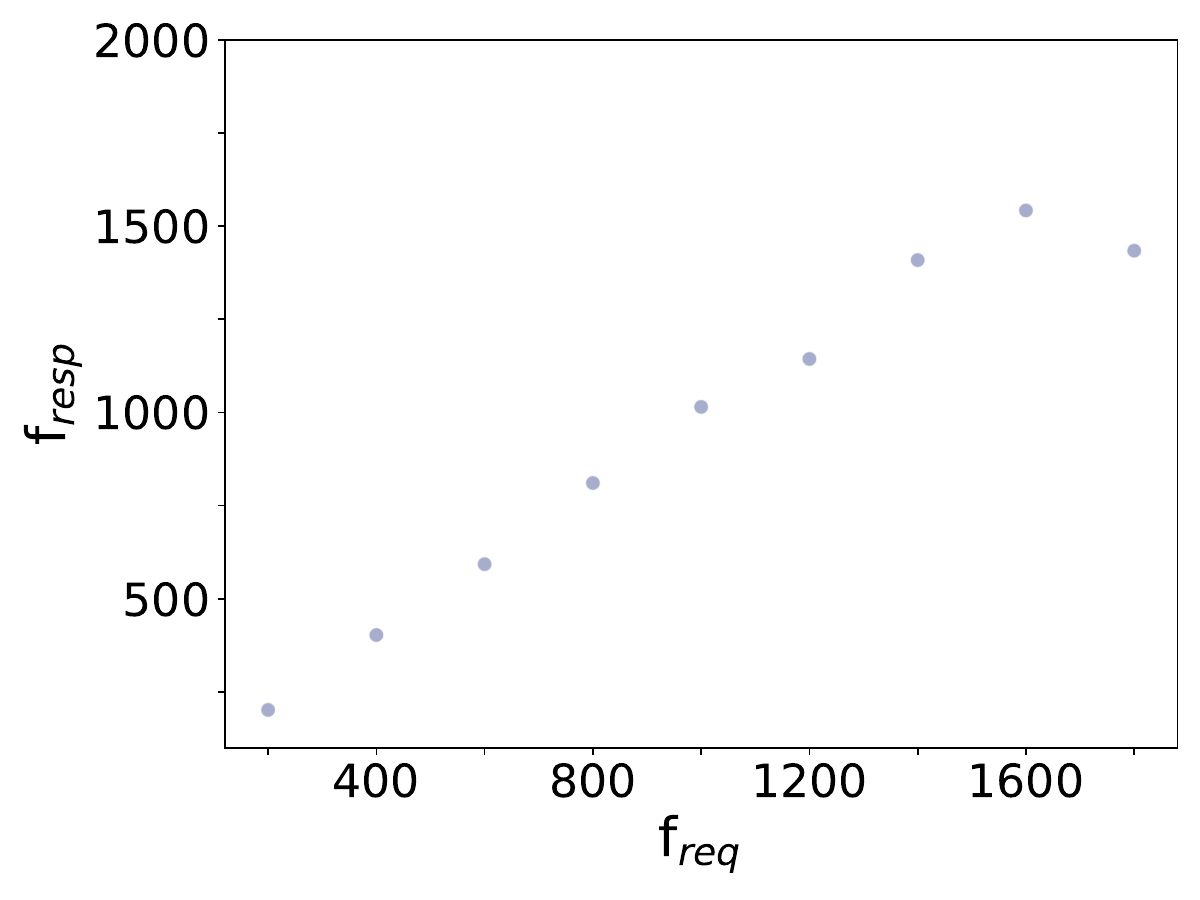}}}

\caption{Fabric -- Rolling mean of throughput with different window sizes.}
\label{fig:throughput_windows}
\end{figure}

In the second part of our analysis, we attempt to find correlations between Fabric's throughput and the components highlighted as potential bottlenecks in the first part, namely peer \ac{CPU} utilization and peer network traffic. 
We start by gaining an overview of how the request rate affects throughput. This is achieved by plotting the throughput as a rolling mean across different window sizes (Figure~\ref{fig:throughput_windows}). A one-second window size results in each data point being plotted individually, revealing how unstable the throughput of this Fabric network gets beyond \ac{freq}=1200\,s$^{-1}$, with fluctuations in \ac{fresp} reaching up to 1000\,s$^{-1}$ within a single run. To see the overall network performance trend, we increase the window size, taking 3~cases at 3~seconds, 8~seconds, and full run. In these cases, the mean is calculated over more data points, and outliers are smoothed out. Here, we see that, on average, the system keeps up with the request rate until reaching approximately \ac{freq}=1600\,s$^{-1}$. Beyond this point, the interrelationship breaks, marking this as the peak throughput observed in the experiment.

\begin{figure}
    \centering
    \includegraphics[width=\linewidth]{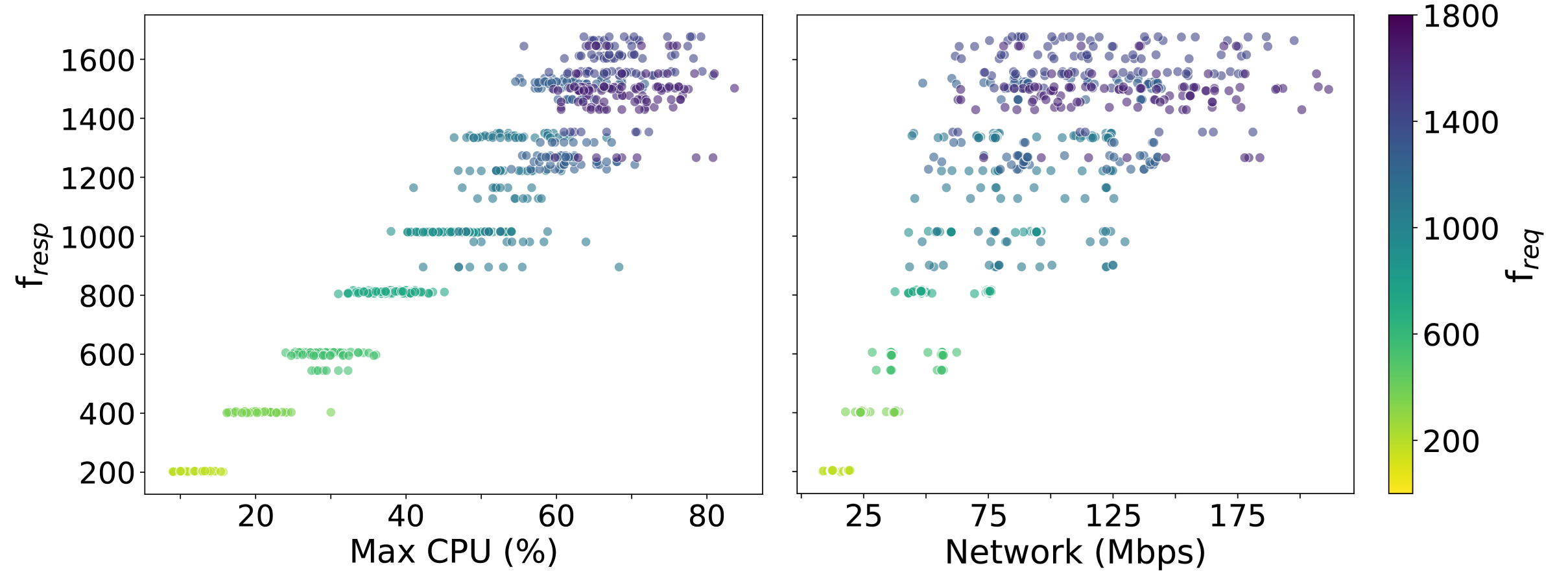} 

    \caption{Fabric -- Throughput against key resources.}
    \label{fig:throughput vs CPU vs Network}
\end{figure}

To obtain insights into the link between peer \ac{CPU} utilization and network traffic, we examine Figure~\ref{fig:throughput vs CPU vs Network}, which plots the two resources against the network's throughput employing a rolling average with a 3-second window. We selected this interval as it matches the average duration required for a transaction to be committed to the blockchain under high request rates. This is significant because queuing effects become prominent at elevated \ac{freq}, and opting for a shorter time window could underestimate throughput.

For \ac{CPU} usage, we observe an initial linear increase with throughput until the characteristic plateau is reached. Both variables seem to increase at a similar rate at the beginning, implying that increased throughput proportionally strains the \ac{CPU}. Additionally, we observe that the instability in \ac{CPU} utilization starts at around \ac{fresp}=1200\,s$^{-1}$, which is also the point at which the correlation between \ac{freq} and \ac{fresp} starts breaking down (Figure~\ref{fig:window_1s}). This indicates that \ac{CPU} usage is more closely correlated to throughput than to the request rate. Similarly, although network traffic increases with throughput, it does so at a lower rate and begins to exhibit instability already at around \ac{fresp}=600\,s$^{-1}$, where the throughput still manages to keep up with the request rate. 

\begin{figure}
    \centering
    \includegraphics[width=0.5\linewidth, trim=0cm 0cm 0cm -2.1cm, clip]{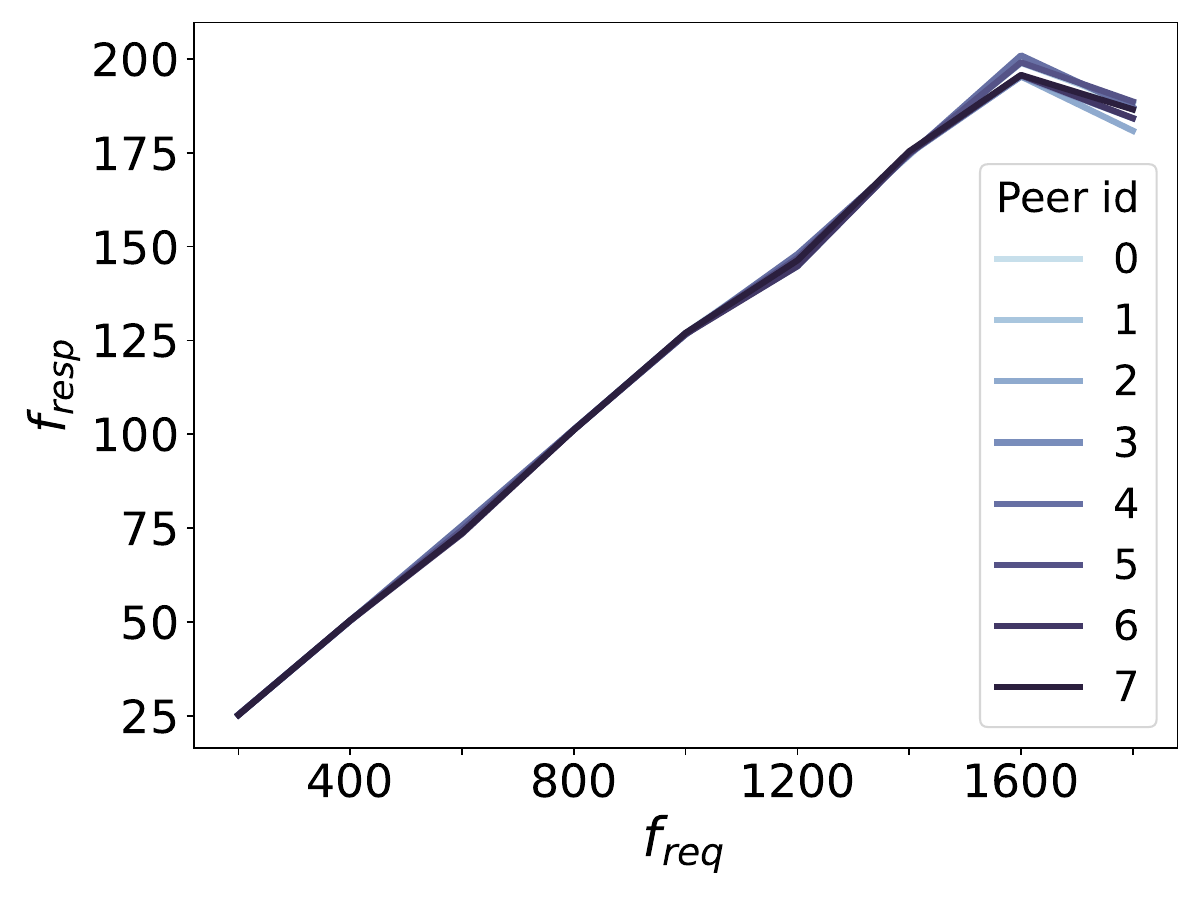} 

    \caption{Fabric -- Throughput per peer.}
    \label{fig:throughput per peer}
\end{figure}

Digging deeper, we focus on the throughput of individual peers (Figure~\ref{fig:throughput per peer}). Here, we see that every peer contributes similarly to the overall throughput, with all of them moving in unison with minor variations at high \ac{freq}. Given our findings in Section~\ref{network}, where we noted significant differences in network traffic between gossip leaders and followers, this outcome suggests that network traffic is not a significant factor in determining throughput compared to \ac{CPU} utilization. If it were, we would expect noticeable differences in throughput between gossip leaders and followers. Consequently, this leaves us with peer \ac{CPU} utilization as the primary factor behind throughput leveling off.

The crucial role of peer CPU utilization further supports the findings in Section~\ref{resource_util} that \ac{vscc} and \ac{mvcc} are the main bottlenecks in Fabric as both depend on peer \ac{CPU}. Given that validations in \ac{vscc} are parallelized and we have noted that the mean core utilization is limited and mean \ac{CPU} utilization is similar across all peers, it appears that the main bottleneck is \ac{vscc}. However, the sequential execution of concurrency controls in \ac{mvcc}, combined with the fluctuating utilization per core featured in Figure~\ref{fig:Chart_timeline_peer0 processors freq=1600}, makes it impossible to determine its role as a bottleneck. As a result, while \ac{vscc} appears to be the main culprit, we cannot conclusively confirm this with the available data. This also highlights a shortcoming of the \ac{dlps}' resource utilization-based approach, which was chosen to make it blockchain agnostic. However, as a trade-off for this generalized approach, the resulting data are less specific than they could be. Therefore, further research would require taking into account and evaluating more precise monitoring data, e.g., using SoundCloud’s Prometheus that is incorporated in Fabric~\cite{Fabric-prometheus}.

\subsection{Quorum: Resource utilization}
\label{resource_util_quorum}

\begin{figure}
\centering
\includegraphics[width=1\linewidth]{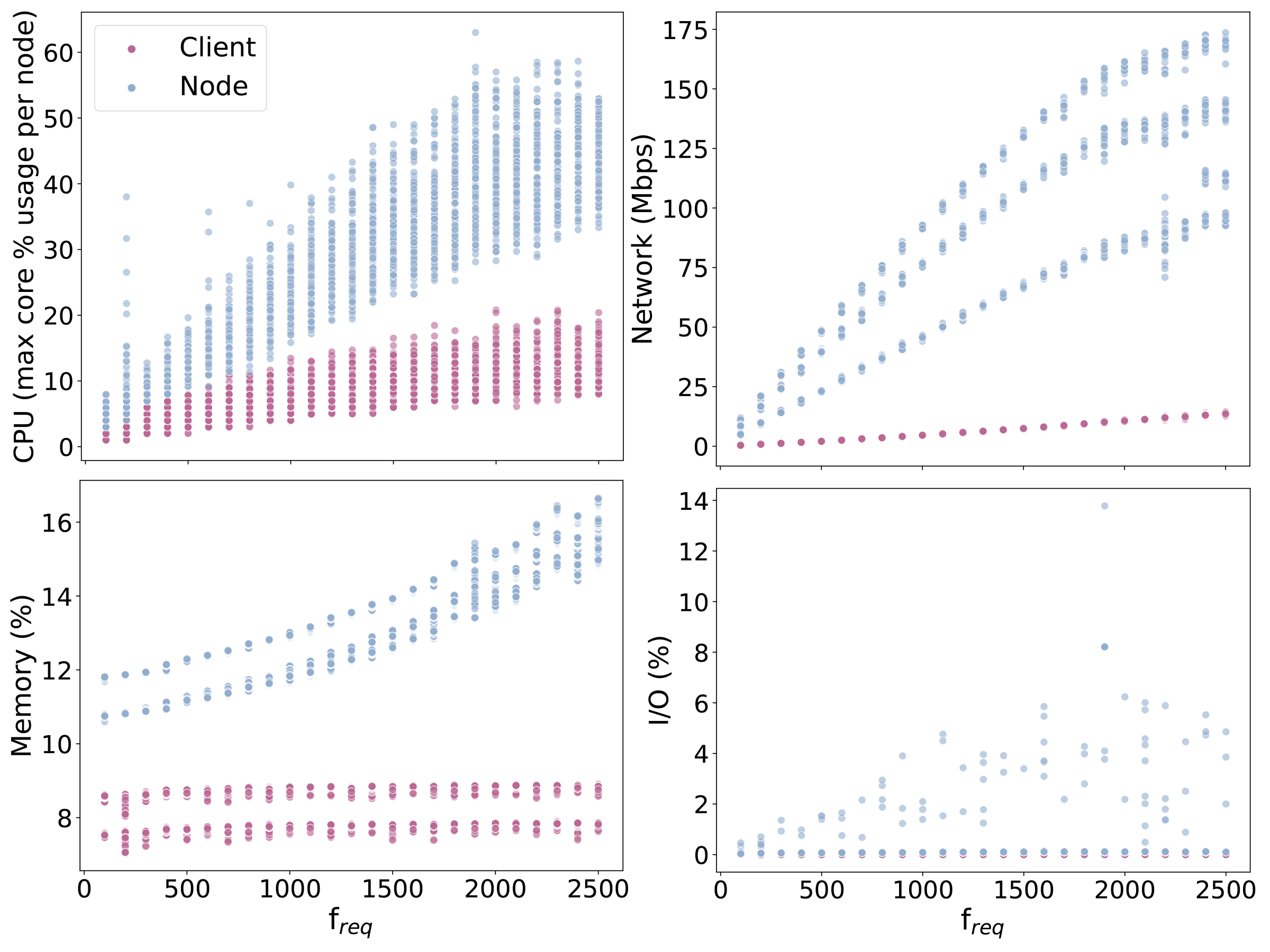}
\caption{Quorum -- Key resource utilization for different request rates \ac{freq}.}
\label{fig:q_Scatter_resources}
\end{figure}

The Quorum analysis, as before, starts by gaining an overview of the four resources in relation to \ac{freq} (Figure~\ref{fig:q_Scatter_resources}). To analyze Quorum's performance in more detail, we refine our analysis by increasing the request rate in increments of 100~requests per second. As with the case of Fabric, even at this early stage, it is apparent that \ac{CPU} and network utilization are closely correlated with the request rate. Memory and hard drive usage are limited, with I/O operations exhibiting high fluctuations but generally registering less than 1~\% usage. Across all resources, client utilization appears to be minimal, and it is either unaffected or grows linearly with the request rate. As a result, in the following sections, we concentrate on the resource utilization of nodes.

\subsubsection{\textbf{CPU}}
\label{q_cpu}

\begin{figure}
\centering
\includegraphics[width=0.7\linewidth]{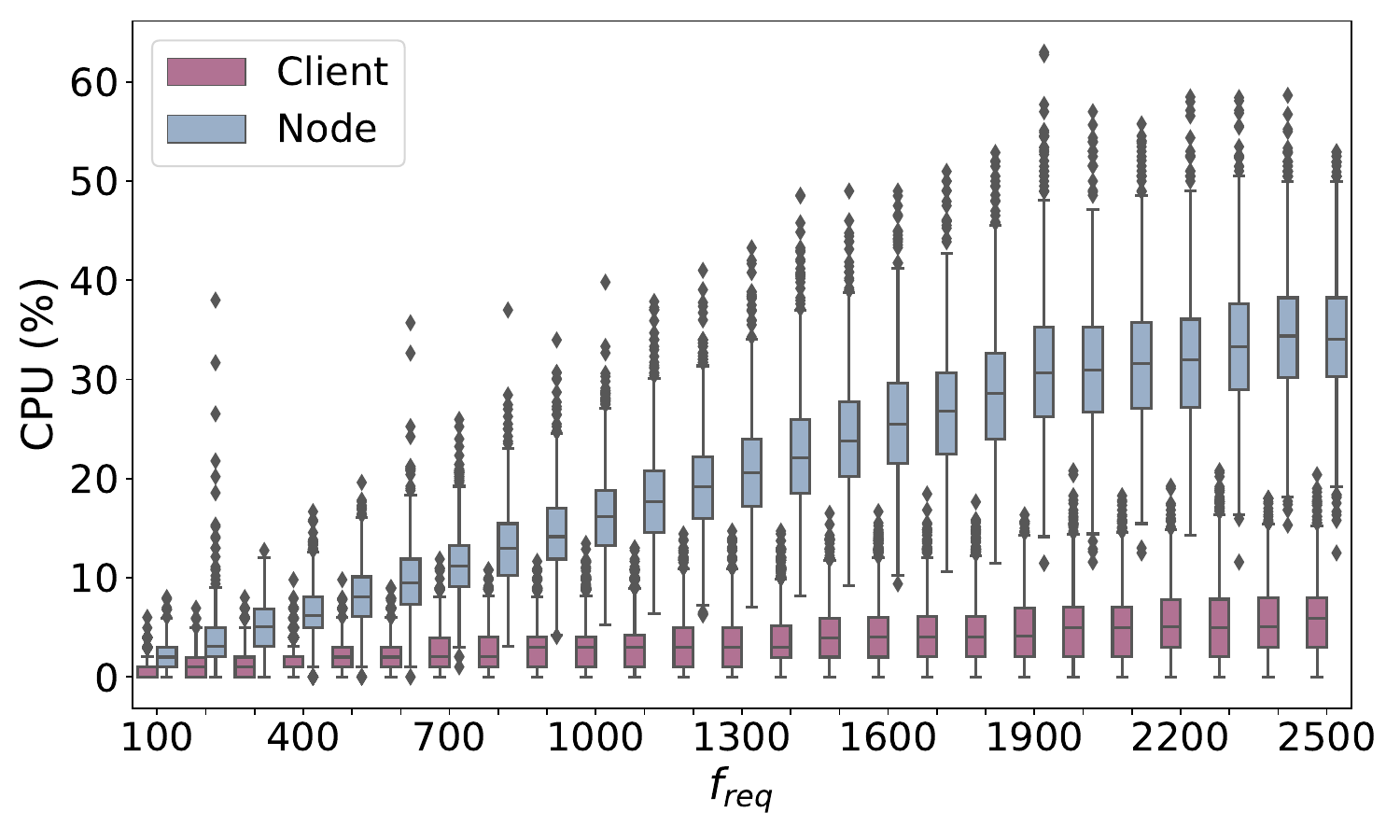}
\caption{Quorum -- CPU utilization of all cores.}
\label{fig:q_cpu_all_cores}
\end{figure}

\begin{figure}
 \subfloat[Mean utilization per Node.\label{fig:cpu_util_per_node}]{{\includegraphics[width=0.5\linewidth]{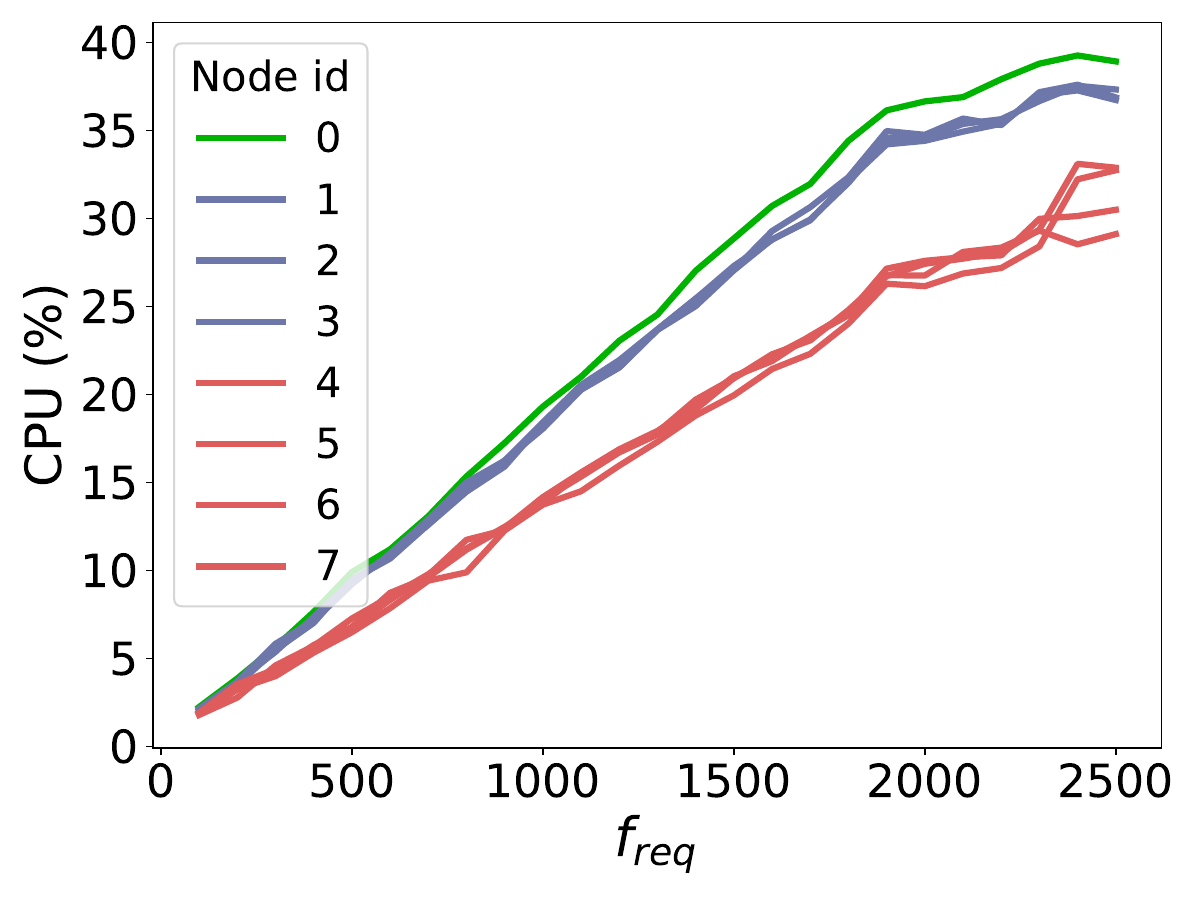} }}
 \subfloat[Mean utilization of Node~0 per core.\label{fig:q_mean_util_core}]{{\includegraphics[width=0.5\linewidth]{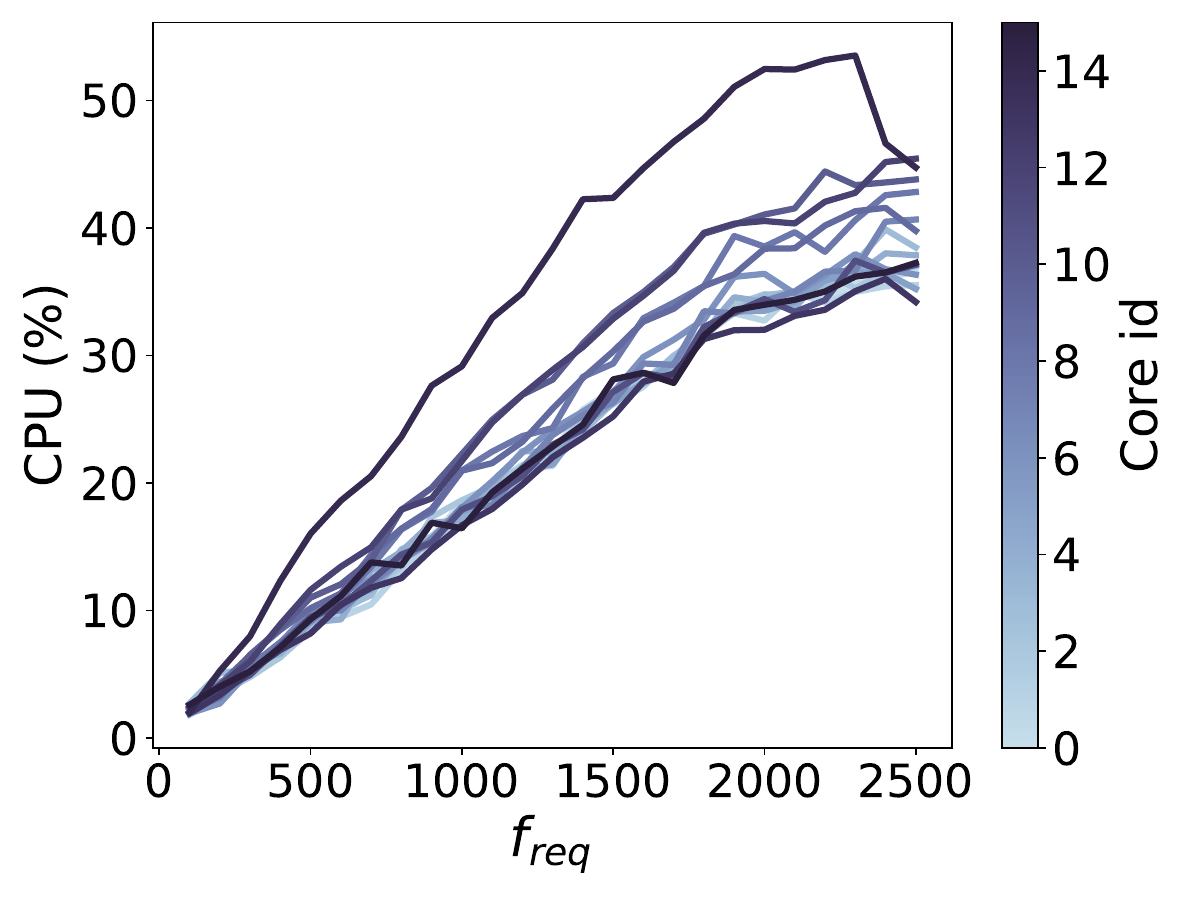}}}
\caption{Quorum -- Mean CPU utilization.}
\label{fig:q_cpu_util}
\end{figure}

We start the \ac{CPU} analysis by looking into the utilization across all cores, where again we note significant fluctuations among the nodes as the request rates increase (Figure~\ref{fig:q_cpu_all_cores}). Examining the mean node \ac{CPU} utilization (Figure~\ref{fig:cpu_util_per_node}), we see that nodes can be grouped into three distinct categories. Node~0~(green) consistently exhibits the highest utilization across all request rates, indicating it is the \ac{raft} leader. This is attributed to the leader's additional operations, such as transaction ordering and compiling transactions into a block. Nodes~1,~2,~and~3~(blue) display slightly lower utilization levels, indicating their role in receiving and pre-validating transactions alongside node~0. In contrast, the remaining nodes (orange) exhibit significantly lower \ac{CPU} usage, with their primary role being the appending of blocks to their local version of the blockchain.

\begin{figure}
 \subfloat[Node~0.
 \label{fig:q_utiil_node0}]
 {{\includegraphics[width=0.5\linewidth]{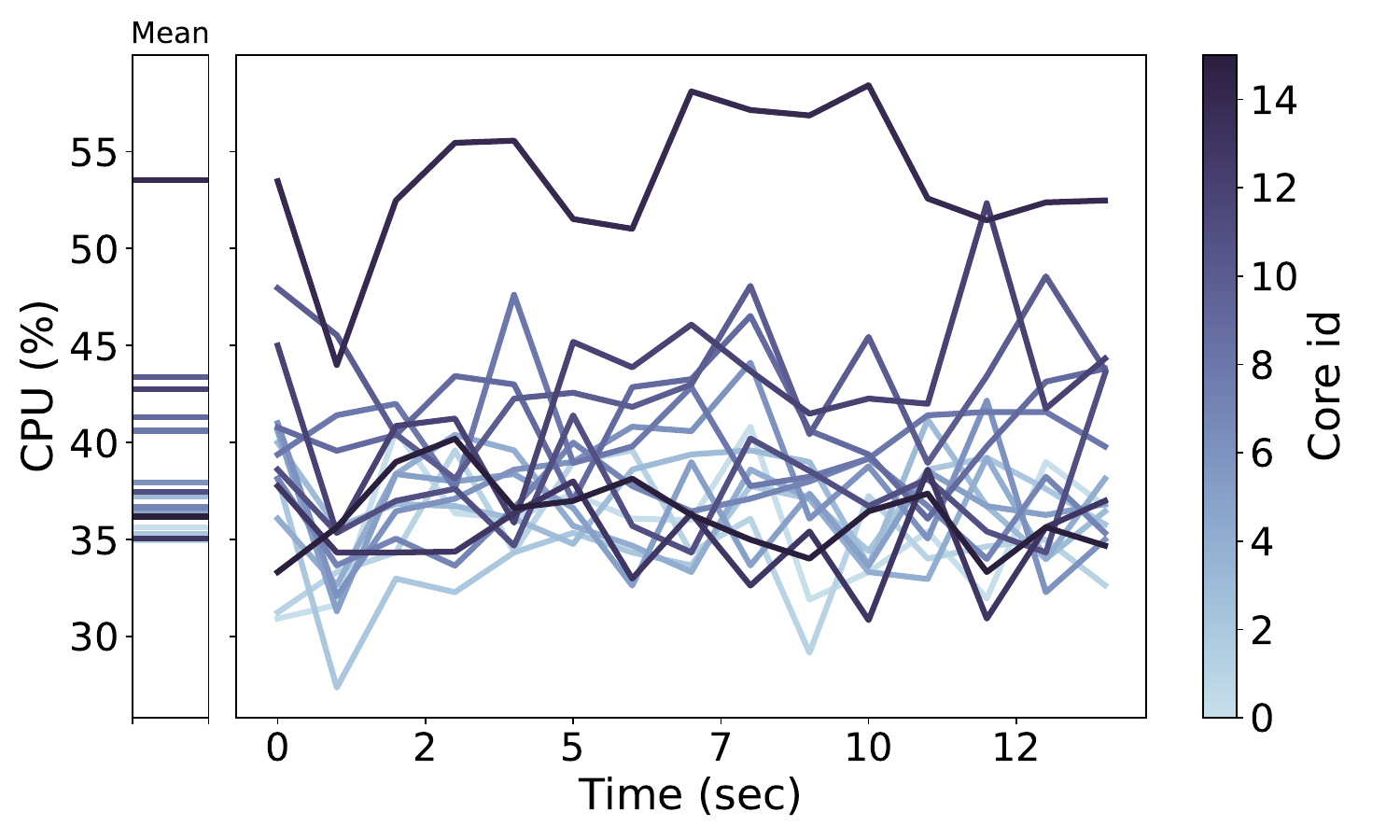}}}
 \subfloat[Node~1.
 \label{fig:q_util_node1}]
 {{\includegraphics[width=0.5\linewidth]{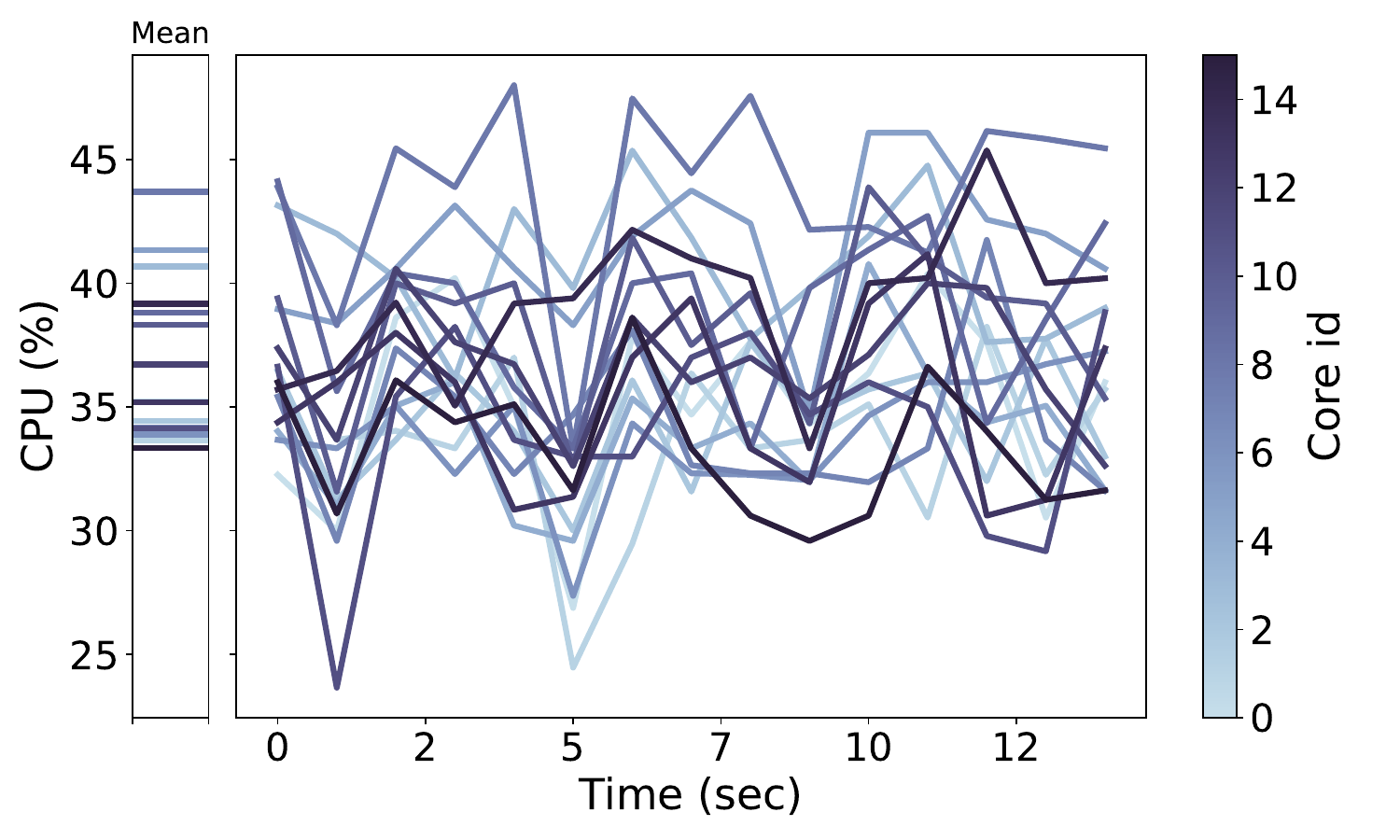}}}
\caption{Quorum -- CPU utilization for \ac{freq}=2300\,s$^{-1}$.}

\label{fig:q_cpu_freq}
\end{figure}

Examining the utilization per core of node~0 (Figure~\ref{fig:q_mean_util_core}) we observe that one core (core~15) bears a higher workload across all request rates. This is due to the fact that with the exception of pre-validation, all other tasks of the leader are executed sequentially, leading to a disproportionate strain on one core. Following this, we examine the temporal evolution of \ac{CPU} usage across individual cores at \ac{freq}=2300\,s$^{-1}$, which is the highest request rate before \ac{CPU} utilization plateaus (Figure~\ref{fig:q_cpu_freq}). Here, we see fluctuations by as much as 20\,\% for both the leader and the follower, with disparities of up to 10\,\% between individual cores, excluding core~15 for node~0. Notably, the main difference between the mean utilization of node~0 and nodes1,~2,~and~3 comes only from core~15. These results suggest that parallel processing in Quorum is even more limited than that of Fabric, with each core's average usage not exceeding 40\,\% and specific tasks of the leader burdening only one core.

With respect to bottleneck detection, nodes~0,~1,~2,~and~3 reach an apparent plateau at \ac{freq}=2400\,s$^{-1}$, while for the remaining nodes, some of them plateau while others do not. Additionally, we observe that there is a considerable decrease in \ac{CPU} usage at \ac{freq}=1800\,s$^{-1}$, which, even though it does not necessarily indicate a bottleneck, could provide clues for identifying factors contributing to the decline in \ac{CPU} utilization. 
From Figure~\ref{fig:cpu_util_per_node}, we observe that node~0 and nodes~1,~2,~and~3 show similar behavior after reaching a plateau. This similarity suggests that transaction ordering and block building, which are the main unique operations performed by the leader, are likely not the sources of the bottleneck. If they were, we would expect the leader's utilization pattern to diverge from that of the other nodes after the plateau. Examining Figure~\ref{fig:q_mean_util_core}, we see a rapid decline in the utilization of core~15 at high request rates. Considering that the remaining sequential operations, such as ordering the transactions, committing the block to the chain, and propagating it through the nodes, typically require minimal \ac{CPU} resources, this sharp drop cannot be justified by these processes, and it appears that another component limits \ac{CPU} utilization. 

\subsubsection{\textbf{Network}}
\label{q_network}

\begin{figure}
\centering
\includegraphics[width=0.5\linewidth, trim=0cm 0cm 0cm -1.6cm, clip]{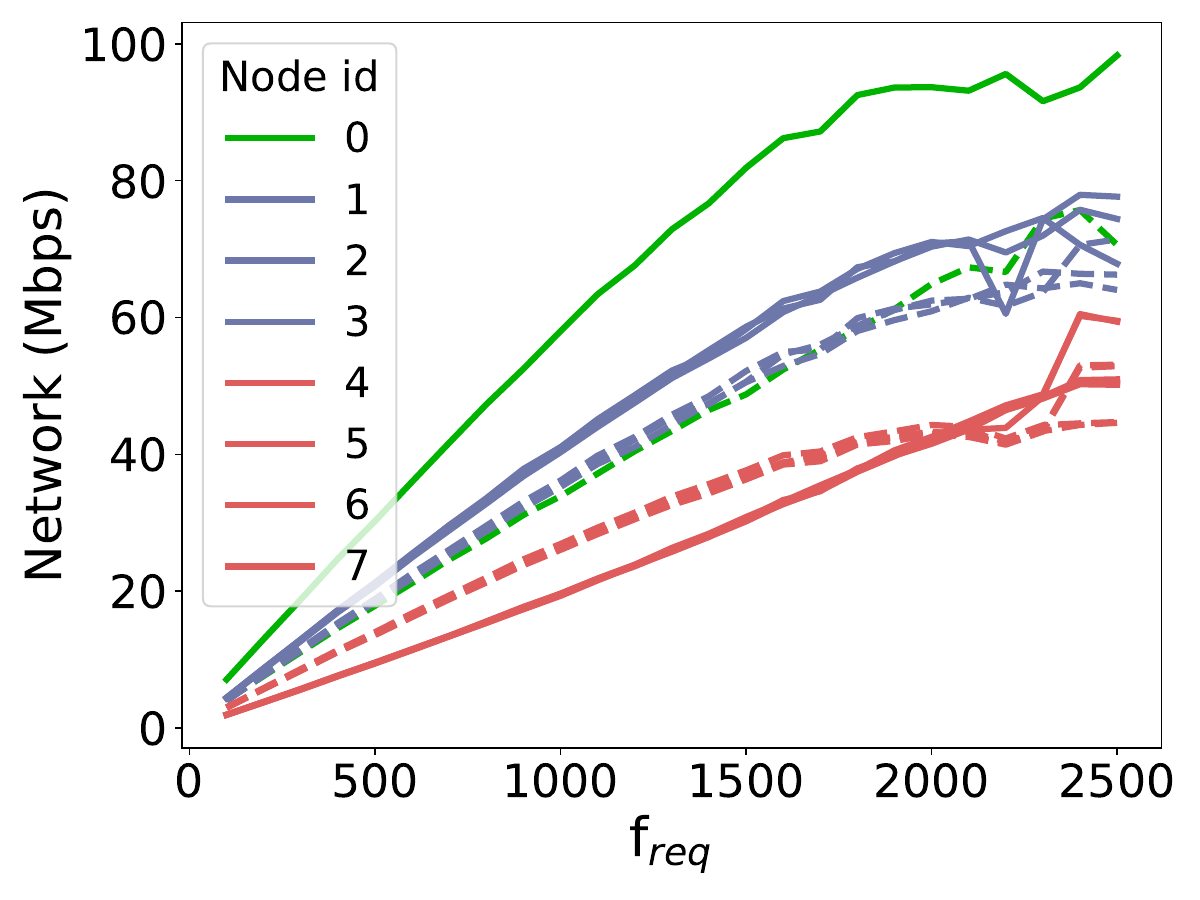}
\caption{Quorum -- Node mean network utilization (outbound traffic as continuous and inbound traffic as dashed lines).}
\label{fig:q_network_nodes}
\end{figure}

Continuing the analysis by looking into the mean network utilization of the nodes (Figure~\ref{fig:q_network_nodes}), we can categorize the nodes into three groups as before. Due to the complexity of the Quorum network traffic, we are not able to decompose the traffic into individual components with adequate accuracy. Therefore, we rely entirely on the architectural design to identify the reasons behind the plateau. As with the case of Fabric, the network analysis does not take into consideration the traffic generated by consensus messages since they generate minimal traffic.

Starting with the blue nodes, we note that their outbound traffic levels off at \ac{freq}=2400\,s$^{-1}$ while signs of plateauing in their inbound traffic appear at \ac{freq}=1800\,s$^{-1}$. The inbound traffic primarily consists of transactions received either directly from clients or through gossip and blocks from the leader, whereas their outbound traffic originates from the dissemination of pre-validated transactions. This suggests that these operations are the limiting factors at their respective request rates.
While we see similar patterns for the orange nodes, the patterns of the leader are essentially the opposite of those of other nodes. The inbound traffic is comprised of all the transactions that are broadcasted to the network and reach the leader through gossip or directly from the clients, and plateaus at \ac{freq}=2400\,s$^{-1}$. The outbound traffic involves mainly block dissemination to the other nodes and plateaus at \ac{freq}=1800\,s$^{-1}$. Since their inbound traffic plateaus at \ac{freq}=2400\,s$^{-1}$, it indicates that the leader keeps receiving the transactions from the blue nodes normally up until that point, leaving us only with block dissemination as the main bottleneck at \ac{freq}=1800\,s$^{-1}$ and transaction propagation as the main issue for \ac{freq}=2400\,s$^{-1}$.

Combining our findings with \ac{CPU} utilization findings, the rapid drop in \ac{CPU} utilization for core~15 at \ac{freq}=2400\,s$^{-1}$ (Figure~\ref{fig:q_mean_util_core}) is caused because the leader does not receive enough transactions from the other nodes and not because the \ac{CPU} cannot keep up with the processes. At \ac{freq}=1800\,s$^{-1}$, the decline could be attributed to either the block propagation or one of the processes preceding it, such as transaction pre-validation and adding the block to the chain, since we already excluded transaction ordering and block building as potential bottlenecks in Section~\ref{q_network}. Considering appending the block to the blockchain is not \ac{CPU} intensive, the primary issues likely lie with block propagation or pre-validation. Using the same line of reasoning for the \ac{freq}=2400\,s$^{-1}$ rate, the bottleneck appears to be either transaction propagation, as already identified, or pre-validation, as it is the only operation that precedes propagation.

\subsubsection{\textbf{Memory \& Hard Drive}}
\label{q_m&h}

\begin{figure}
 \subfloat[Memory utilization.\label{fig:q_memory}]{{\includegraphics[width=0.5\linewidth]{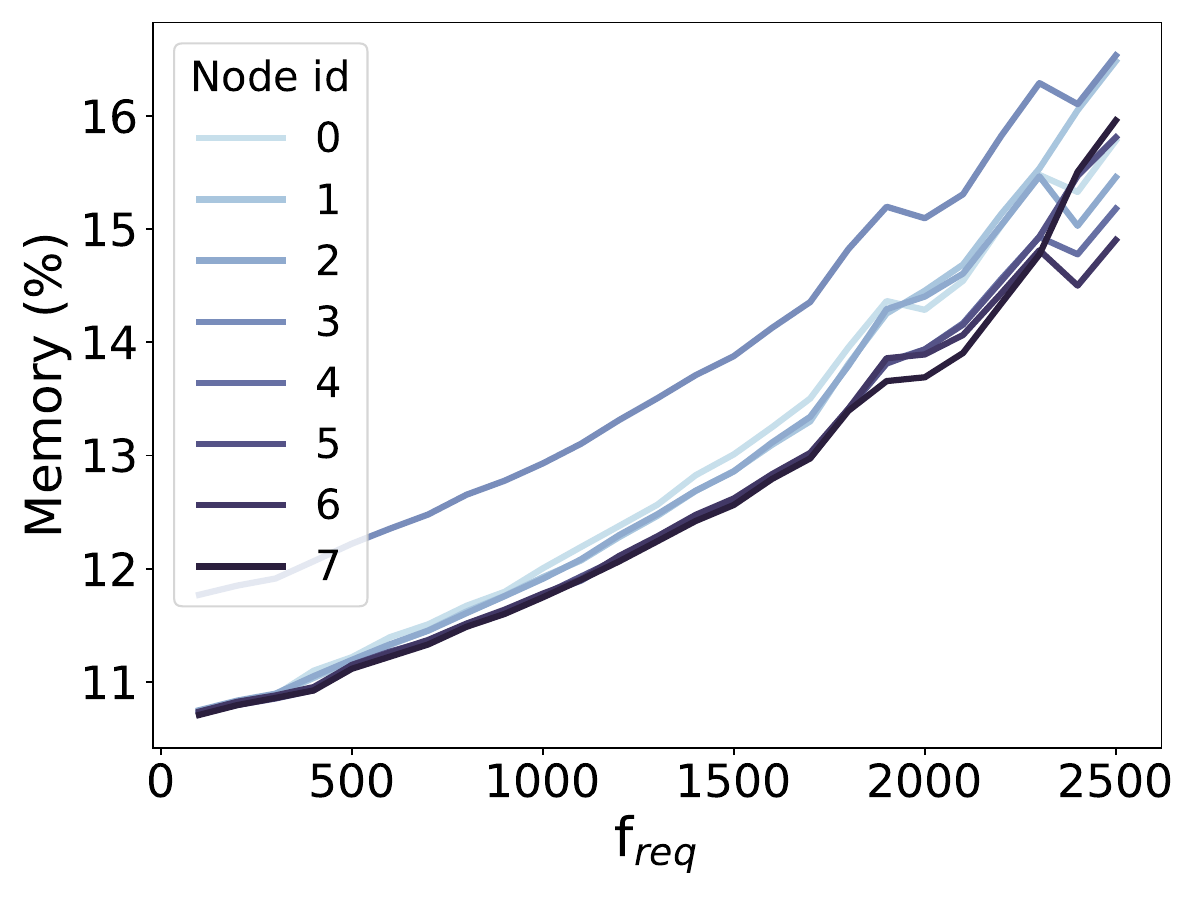} }}
 \subfloat[Hard drive utilization.\label{fig:q_io}]{{\includegraphics[width=0.5\linewidth]{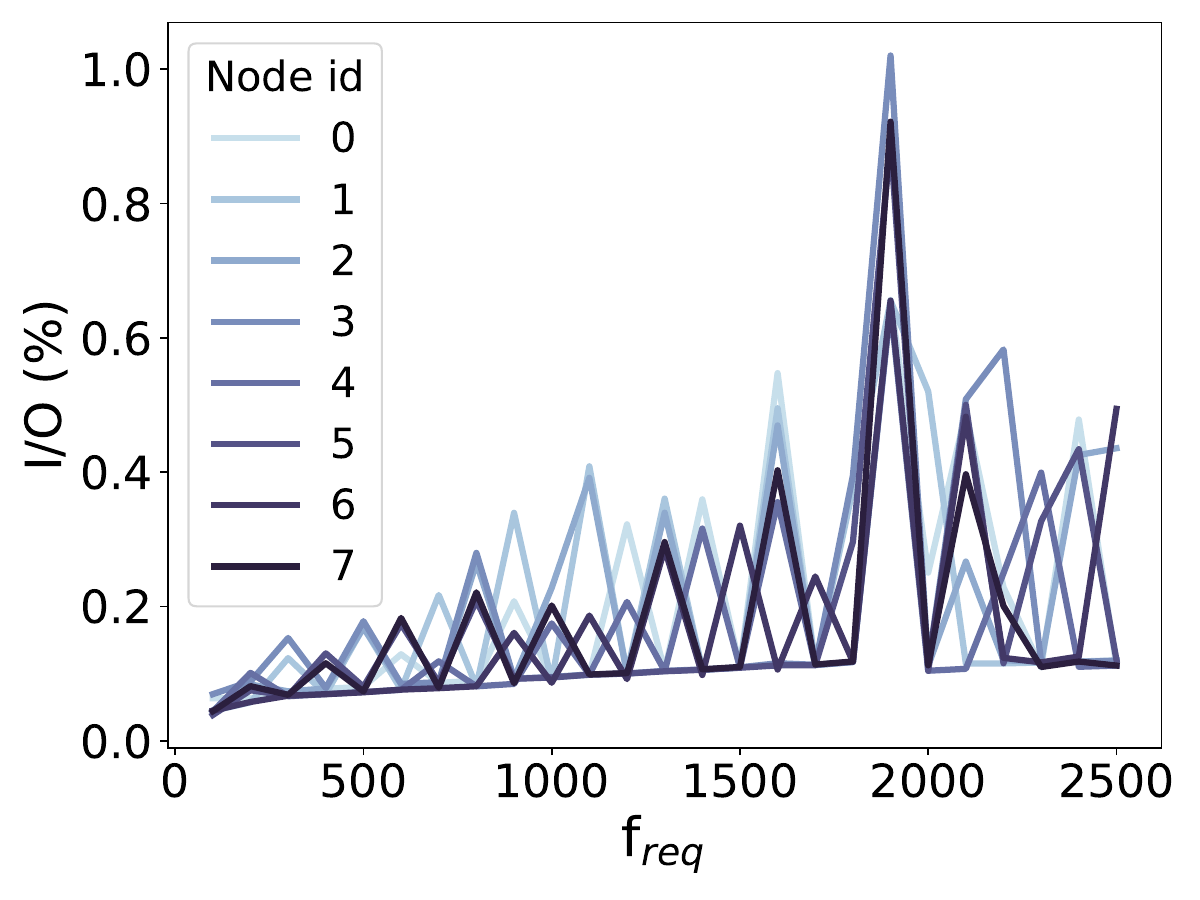}}}
\caption{Quorum -- Mean utilization per node.}
\label{fig:q_memory_util}
\end{figure}

Similar to our observations with Fabric, Figure~\ref{fig:q_Scatter_resources} indicates that neither memory nor hard drive utilization are likely to be bottlenecks. Nevertheless, we proceed to examine the utilization metrics of the nodes. The analysis reveals consistent behavior in memory utilization across all nodes, with no signs of plateauing (Figure~\ref{fig:q_memory}). This uniformity suggests that memory is not contributing to performance degradation. Despite observing significant peaks in hard drive utilization in Figure~\ref{fig:q_Scatter_resources}, the mean utilization remains below 1\,\% (Figure~\ref{fig:q_io}), indicating it is not a constraining factor either. The large fluctuations are probably related to the writing of the block into each node's database, but since it is improbable that it leads to a bottleneck, we desist from further investigations.

Concluding the first part of the Quorum analysis, we identify two primary points where there is a significant drop in performance. The first one appears at \ac{freq}=1800\,s$^{-1}$, with the main factors likely being block propagation or the pre-validation of transactions. The second one is observed at \ac{freq}=2400\,s$^{-1}$, which appears to stem from the propagation of transactions through the network or the pre-validation of transactions.

\subsection{Quorum: Throughput}
\label{q_throughput}

The throughput analysis of Quorum starts by examining its correlation with the request rate. We see that even for small window sizes, throughput remains relatively stable but starts to exhibit higher fluctuations at \ac{freq}=1800\,s$^{-1}$. Examining the full run, we observe that throughput plateaus at around \ac{fresp}=2100\,s$^{-1}$, which is significantly lower than the maximum request rate. This suggests that the overall performance of the blockchain started to decline long before reaching the highest request rate. This observation suggests that the performance degradation noted at \ac{freq}=1800\,s$^{-1}$ for both \ac{CPU} and network utilization may be more critical in identifying the bottleneck.

\begin{figure}
 \subfloat[1s window\label{fig:q_window_1s}]{{\includegraphics[width=0.5\linewidth]{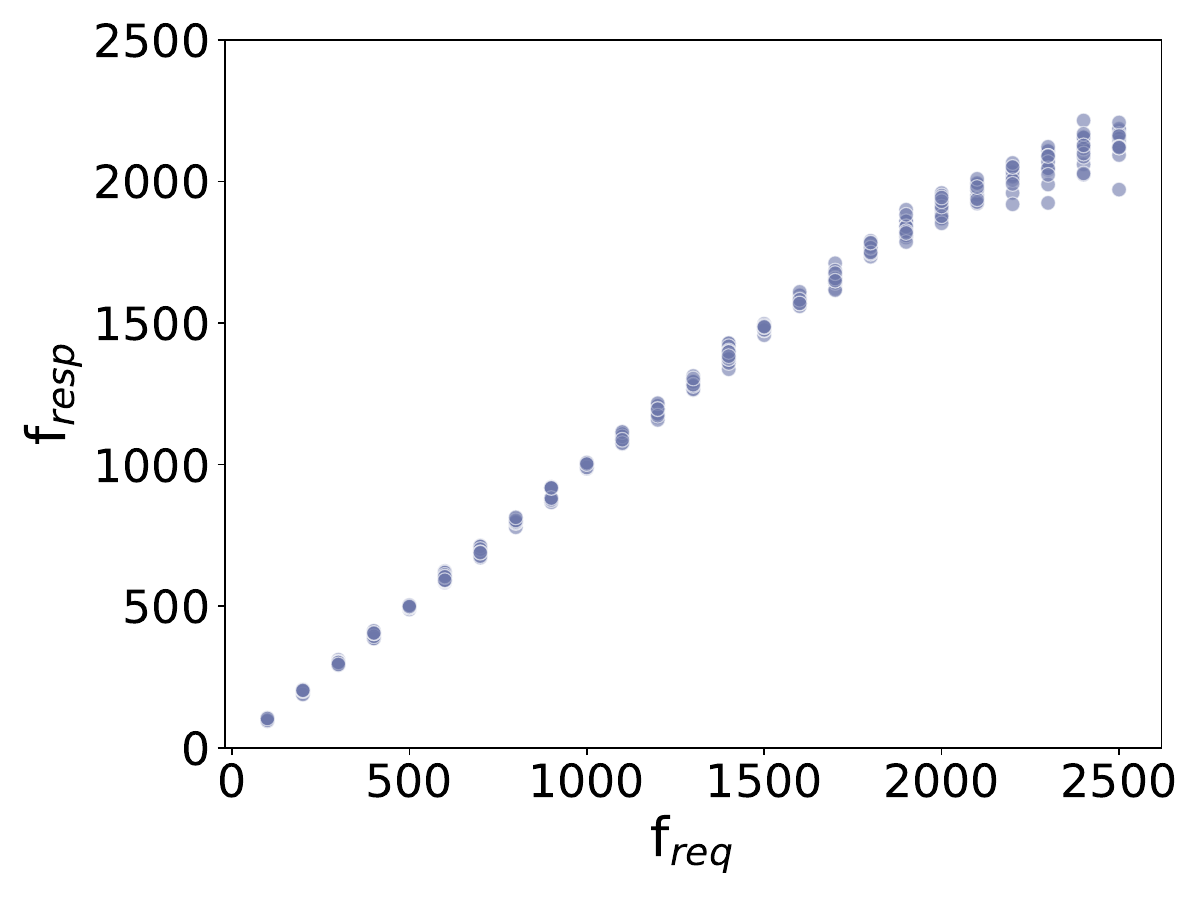} }}
 \subfloat[3s window\label{fig:q_window_3s}]{{\includegraphics[width=0.5\linewidth]{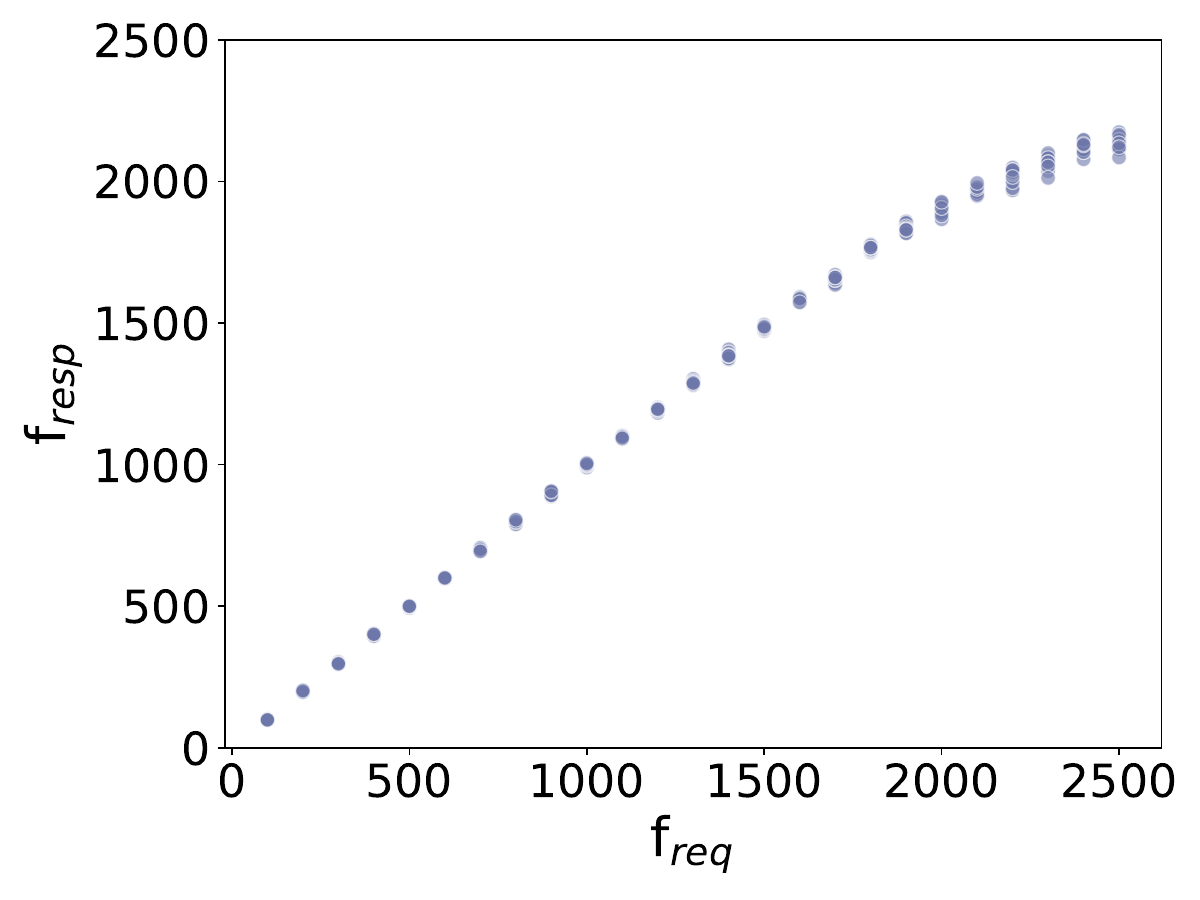}}}

  \subfloat[8s window\label{fig:q_window_8s}]{{\includegraphics[width=0.5\linewidth]{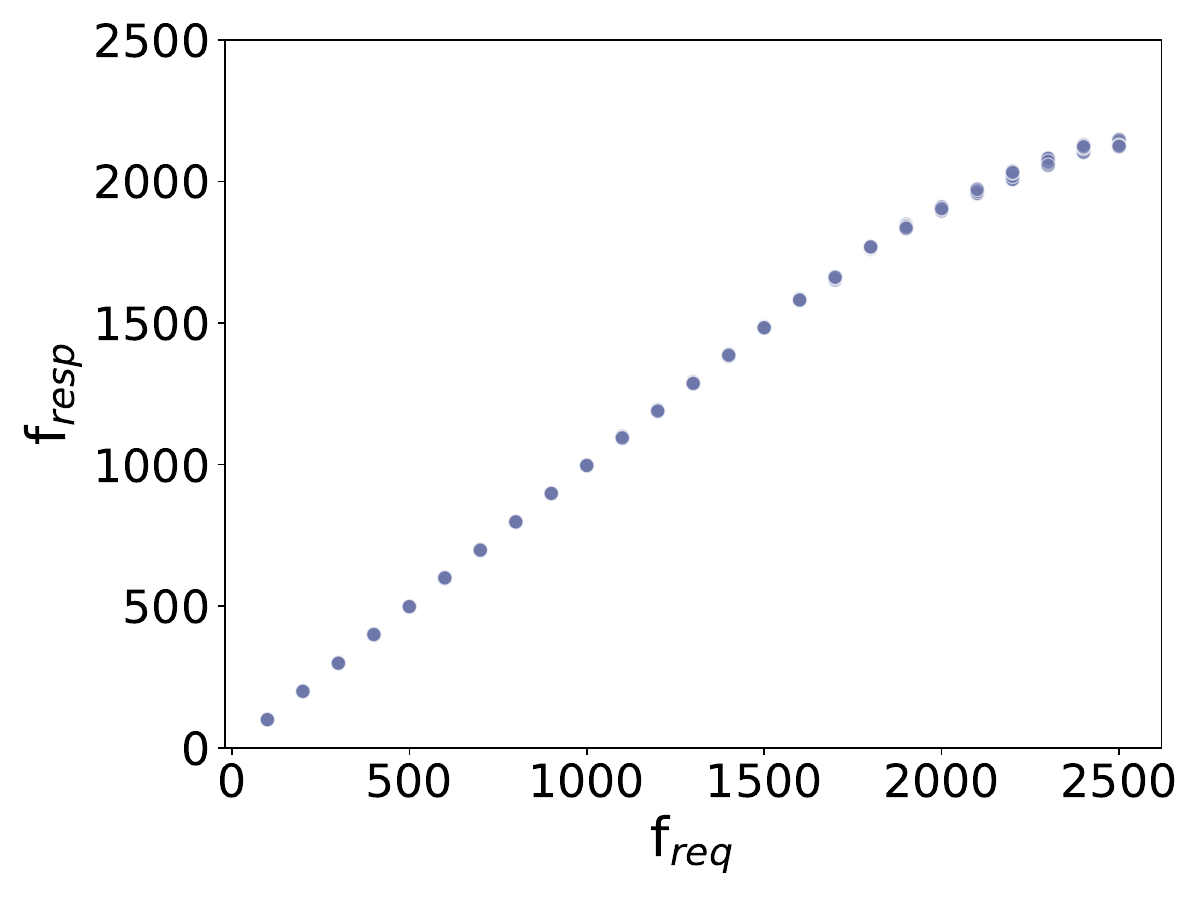} }}
 \subfloat[15s window\label{fig:q_window_14s}]{{\includegraphics[width=0.5\linewidth]{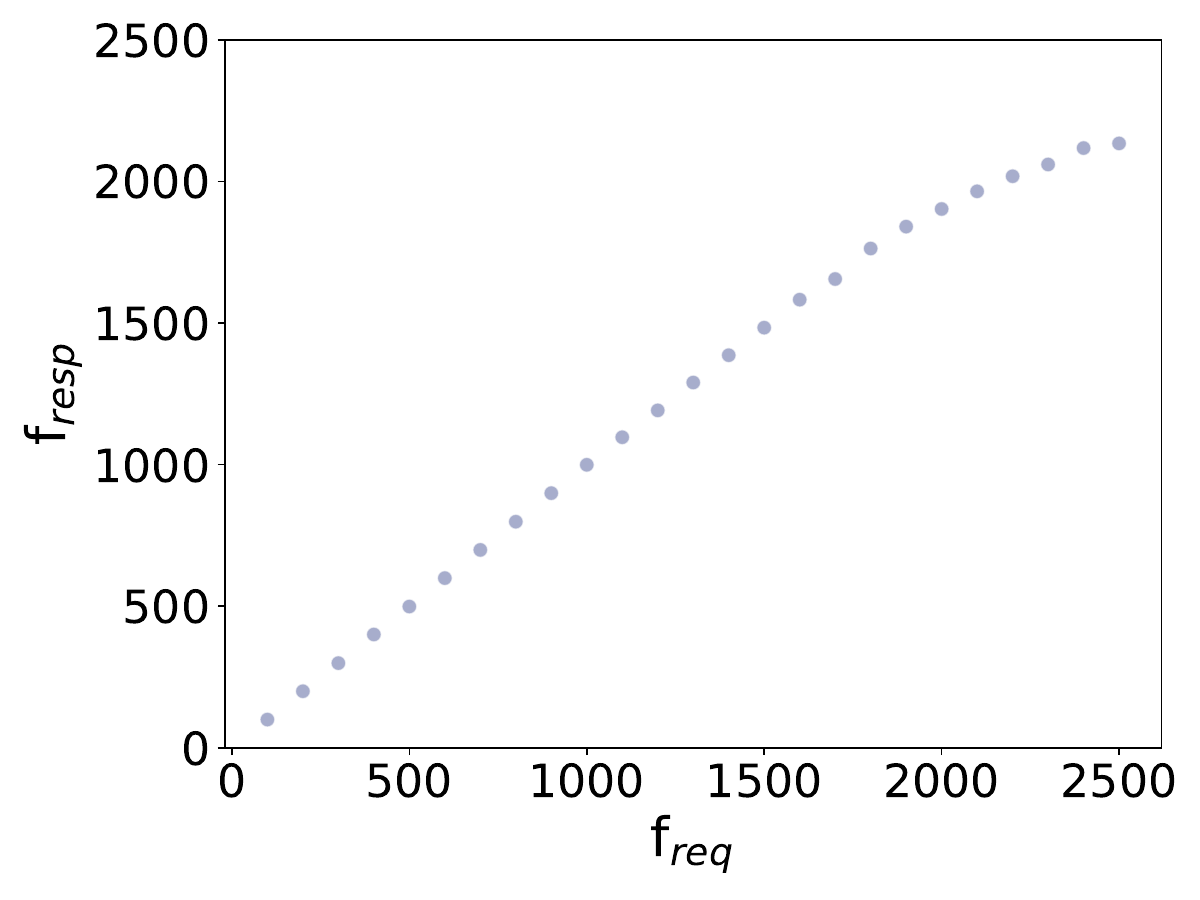}}}

 \caption{Quorum -- Rolling mean of throughput with different window sizes.}
\label{fig:q_throughput_windows}
\end{figure}

\begin{figure}
    \centering
    \includegraphics[width=\linewidth]
    {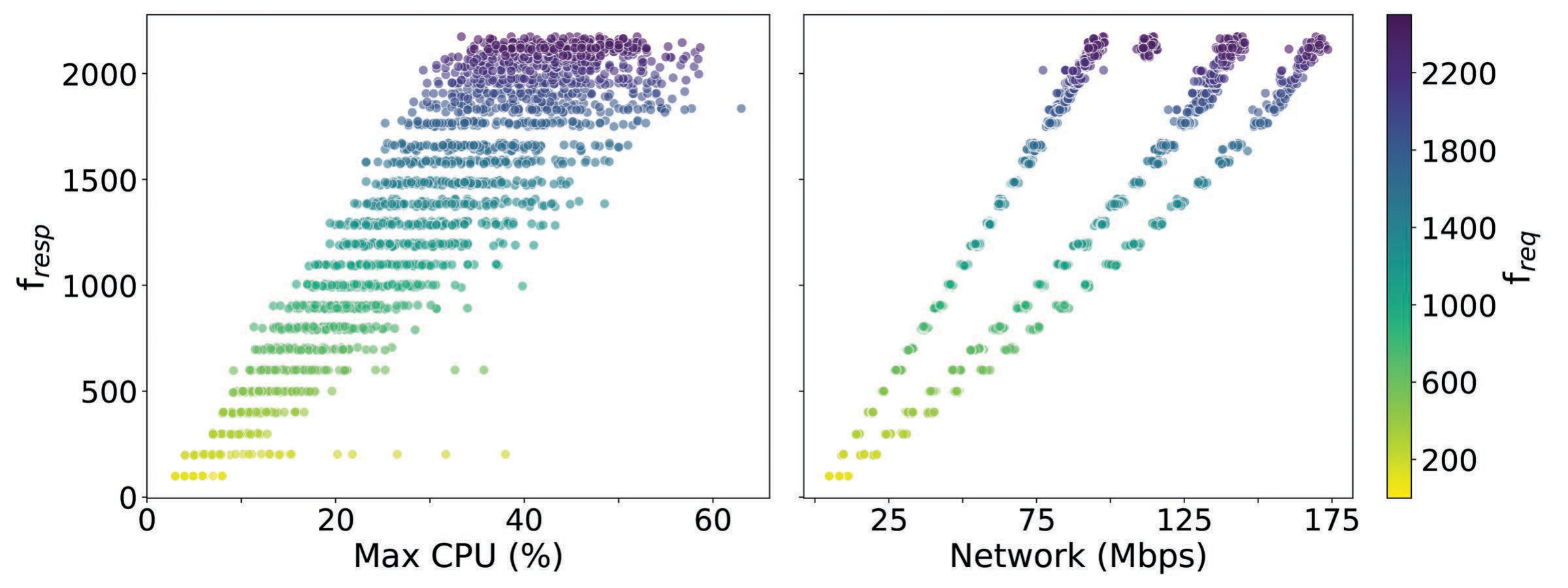} 

    \caption{Quorum -- Throughput against key resources.}
    \label{fig:q throughput vs CPU vs Network}
\end{figure}

Examining throughput against the \ac{CPU} and network utilization (Figure~\ref{fig:q throughput vs CPU vs Network}), we see similar growth rates and patterns. Both resources keep up with throughput at the beginning and experience higher performance fluctuations at around \ac{fresp}=1800\,s$^{-1}$. Beyond this point, the differences in throughput between request rates become less pronounced, becoming indistinguishable as it approaches \ac{fresp}=2100\,s$^{-1}$.

Since both resources exhibit similar patterns, we search for a common source behind the blockchain's performance degradation. Considering that block and validated transaction propagation are network-related, if they were the bottleneck, they would mainly impact network traffic. As a result, they are less likely to be the bottleneck. This leaves only transaction pre-validation as the possible constraining factor, which can impact both \ac{CPU} and network as it represents the first step in processing a transaction.

\begin{figure}
    \centering
    \includegraphics[width=0.5\linewidth, trim=0cm 0cm 0cm -2cm, clip]{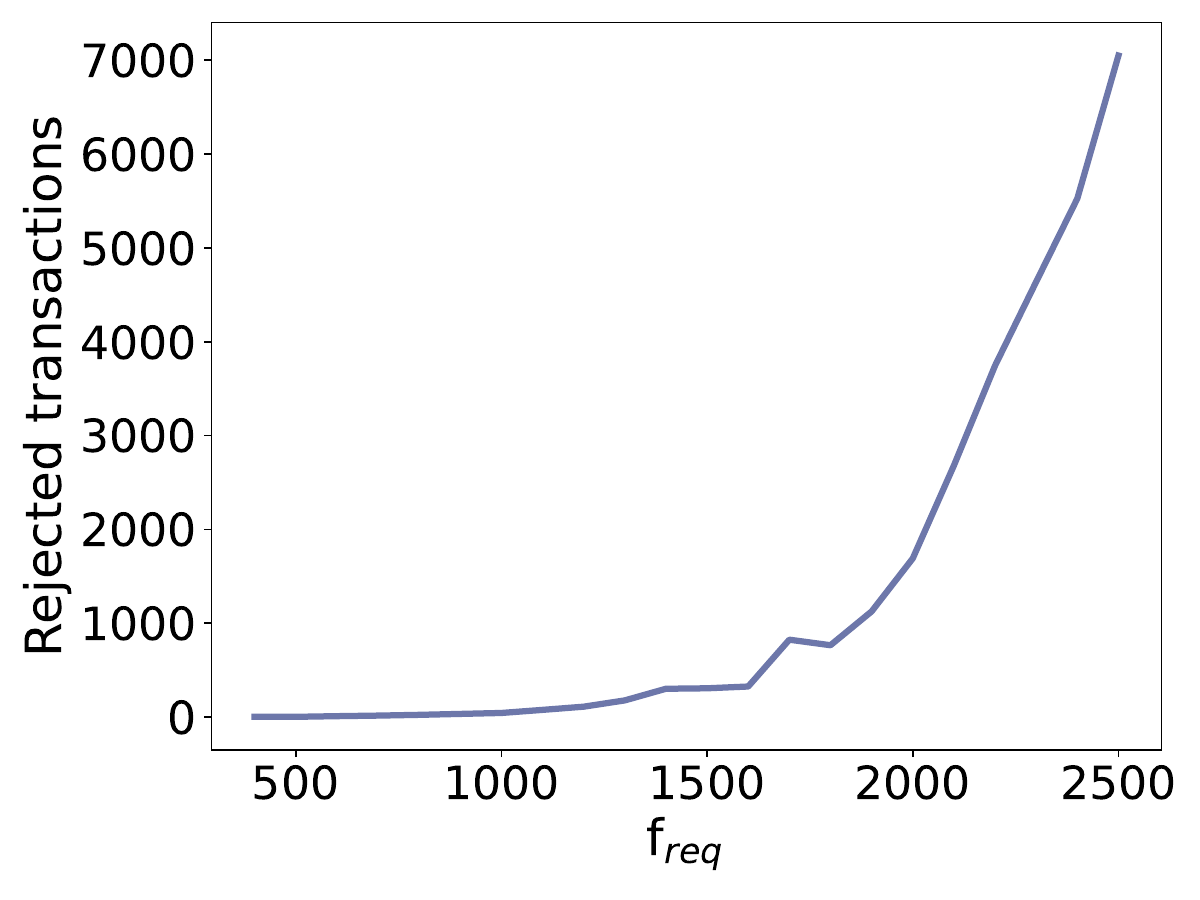} 

    \caption{Quorum -- Mean number of rejected transactions.}
    \label{fig:rejected_txs}
\end{figure}

Looking further into pre-validation, we deduce that the bottleneck is either pre-validation itself, i.e., the nodes do not validate the incoming transactions fast enough, which in turn slows down the network, or the nodes do not receive enough transactions. From  Figure~\ref{fig:q_Scatter_resources}, we see that the network utilization of clients increases linearly with the request rate, which suggests that they send the proper number of transactions to the nodes. 
Examining further the interaction of nodes with the incoming transactions, we look into the number of rejected transactions (Figure~\ref{fig:rejected_txs}). By the term rejected transactions, we refer to those transactions that nodes decline to propagate through the network, resulting in clients receiving almost immediate (within less than 50ms) notification of transaction failure after submission. Initially, the count of rejected transactions remains minimal but begins to surge at \ac{freq}=1800\,s$^{-1}$, culminating in approximately 7000 rejections by \ac{freq}=2500\,s$^{-1}$. This escalation translates to a rejection rate of 14\,\%  before reaching a plateau. This significant increase in rejections prompts us to investigate further to understand the underlying causes.

\begin{figure}
 \subfloat[Cumulative number of transa-\\ctions in the pool.
 \label{fig:q_full_pool}]
 {{\includegraphics[width=0.5\linewidth]{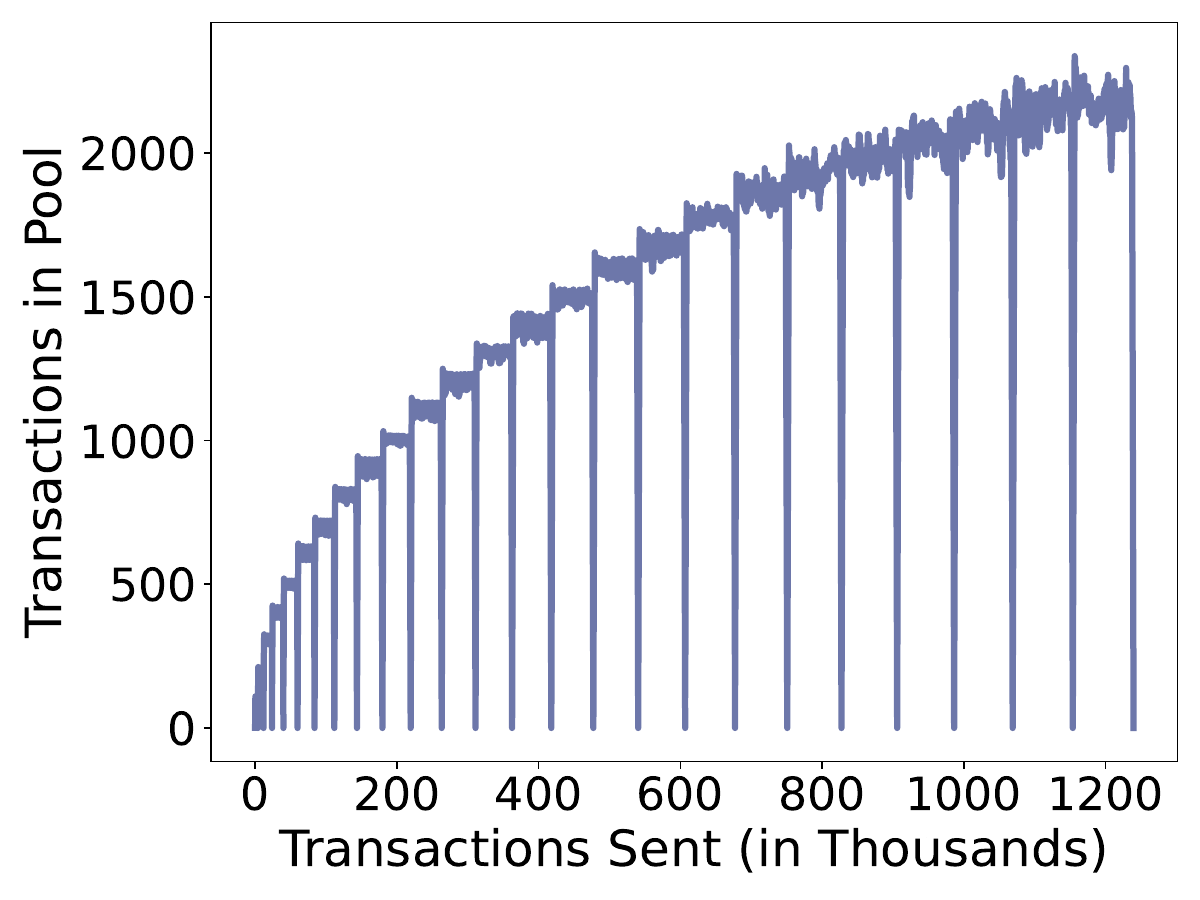} }}
 \subfloat[Number of transactions in the\\ pool per client.
 \label{fig:q_client_pool}]
 {{\includegraphics[width=0.5\linewidth]{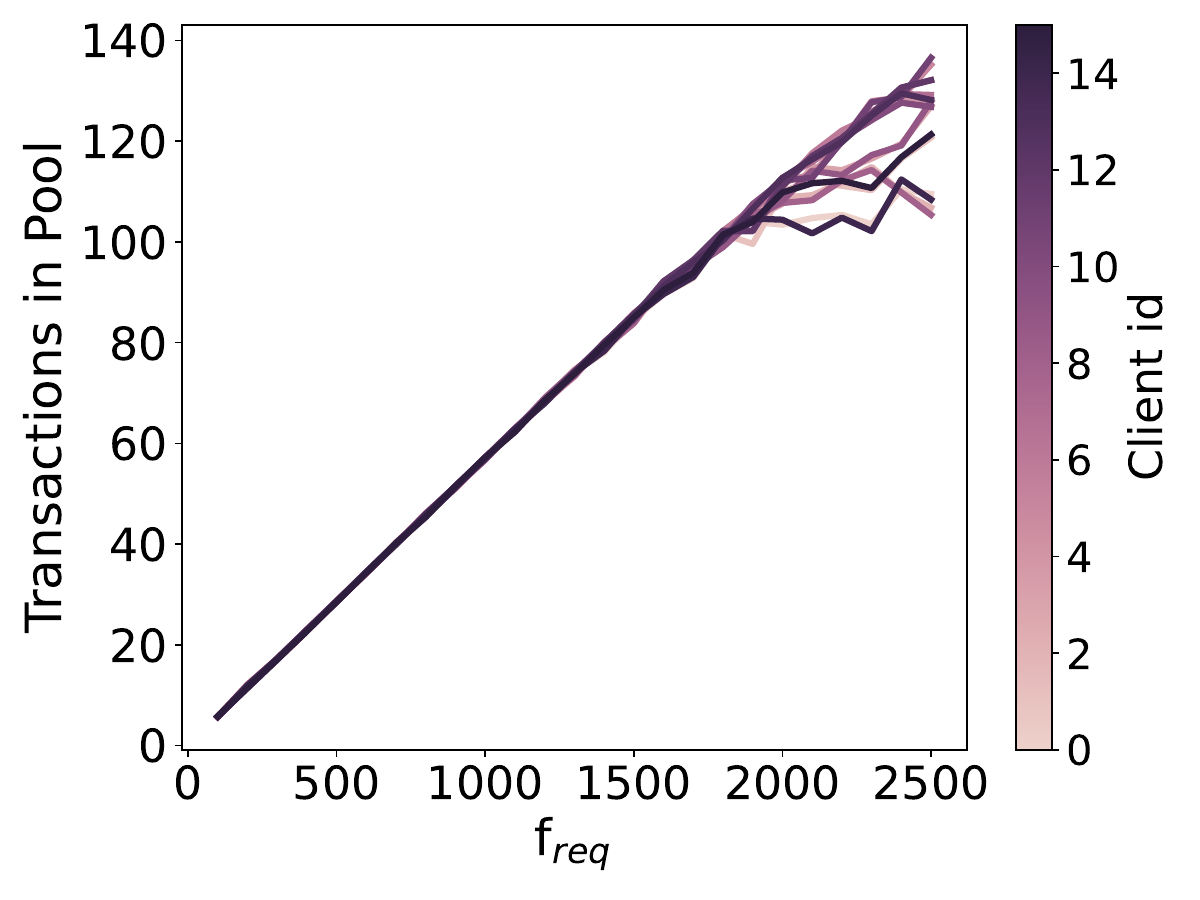}}}
\caption{Quorum -- Transactions in the pool.}
\label{fig:q_pool}
\end{figure}

In Quorum, transactions are categorized as either executable or non-executable. Executable transactions are the ones that can be immediately included in a block, while non-executable transactions are out of nonce order and must wait for preceding transactions with lower nonce to execute first. Investigating the reasons behind the increasing rejection rates, we identify two main constraints in Quorum's transaction handling. First, Quorum caps the transaction pool at 4096~executable and 10000~non-executable transactions~\cite{noauthor_quorumcmdutilsflagsgo_nodate}. However, according to Figure~\ref{fig:q_full_pool}, the transaction pool's capacity is never fully utilized, suggesting that this is not the cause behind the rejections. The periodic drop in the number of transactions in the pool aligns with the end of each ramp-up phase of DLPS, where nodes commit pending transactions to blocks and clear the pool before the new ramp-up commences.

The second limitation involves the clients, who are limited to 16~executable and 500~non-executable transactions in the pool at any given time~\cite{noauthor_quorumcmdutilsflagsgo_nodate}. Examining Figure~\ref{fig:q_client_pool}, we see that at \ac{freq}=1800\,s$^{-1}$, transaction submission by six clients begins to exhibit instability, diverging from the previous linear relationship with the request rate, and most clients diverge at \ac{freq}=2400\,s$^{-1}$. This pattern suggests that clients are reaching the limit of executable transactions in the pool, leading to an increased rate of rejections. To answer this definitively, we would have to separate the client transactions into executable and non-executable and examine which type is getting rejected, something that is not possible due to \ac{dlps}' blockchain agnostic nature that does not allow us to distinguish transactions between the different types. However, considering that the number of rejected transactions increases sharply at \ac{freq}=1800\,s$^{-1}$, combined with the observed reduction in \ac{CPU} and network utilization starting at the same point and the subsequent plateau at \ac{freq}=2400\,s$^{-1}$, strongly suggest that the primary bottleneck is the limit on the number of executable transactions per client.

\section{Conclusion}
\label{sec:conclusion}

This paper introduces a general illustrative method for blockchain bottleneck identification, demonstrated through an in-depth analysis of a 12-node Fabric network and an 8-node Quorum network. Our method leverages \ac{eda} to analyze blockchain performance metrics, highlighting their specific characteristics and bottlenecks. By employing a combination of proportional analysis and the study of plateau-shaped trends in resource utilization versus transaction metrics, we uncover anomalies across varying time windows. The approach allows us to draw conclusions by comparing the correlation between data trends, the request rates (\ac{freq}), and response rates (\ac{fresp}).

For the case of Fabric, using our method, we were able to identify that the validation
phase is the main bottleneck in Fabric’s performance, even for~v2.0, with \ac{vscc} being the most likely component behind the bottleneck. We were also able to showcase the average degree of parallelization within Fabric, which leaves ample room for improvement. These findings are in line with previous investigations \citep{thakkar_scaling_2021}. Further studies could also attempt to dive deeper into the validation phase using tools tailored to Fabric to identify with higher confidence the role of \ac{vscc} and \ac{mvcc} for the blockchain's performance bottlenecks.

For Quorum, we determined that the bottleneck stems from the restriction on the number of executable transactions a client can have in the transaction pool, which results in a significant number of rejected transactions. Additionally, our findings illustrate Quorum's relatively limited capacity for parallel processing. Future research could again rely on more specialized tools to examine thoroughly the role of executable and non-executable transactions in the network's performance. 

Our study is not without limitations. Specifically, in the network configurations we studied, all nodes were located in the same data center, while in real-life scenarios, nodes would likely be scattered around the globe. Related work has already analyzed performance degradation with significant network latencies~\cite{guggenberger_-depth_2022}. As a result, our analysis poses the best-case scenario for network conditions, and additional research is needed to evaluate it under real-life situations. Moreover, while our approach offers a streamlined analysis, a deep understanding of blockchain-specific mechanisms remains crucial for accurate diagnoses.

We believe our work, including the publication and the open-source code, provides researchers and practitioners with a clear and illustrative starting point for simpler blockchain performance and bottleneck analysis. It can also serve to analyze other blockchains or as a stepping point for researchers to develop their approaches while offering a deeper understanding of the individual components of the underlying blockchain and their interrelationships.

\begin{acks}
This work was supported in part by the Luxembourg National
Research Fund (FNR) in the FiReSpARX (ref. 14783405) and PABLO (ref. 16326754) projects and by PayPal, PEARL grant (ref. 1334293). Additional funding was provided by the Bavarian Ministry of Economic Affairs, Regional Development and Energy for their funding of the Fraunhofer Blockchain Center project (ref. 20-3066-2-6-14). 
\end{acks}


\bibliographystyle{ACM-Reference-Format}
\bibliography{sample}

\end{document}